\documentclass[twocolumn,times]{aastex61}
\pdfoutput=1 
\usepackage{amsmath,amstext,natbib,float,epsfig,multirow}
\usepackage[figure,figure*]{hypcap}

\newcommand{\lyaf}[1]{Ly$\alpha$ forest}
\newcommand{\lya}[1]{Ly$\alpha$}
\newcommand{\kms}{\ensuremath{\mathrm{km\;s^{-1}}}}
\newcommand{\hMpc}{\ensuremath{h^{-1}\,\mathrm{Mpc}}}
\newcommand{\cMpc}{\ensuremath{\mathrm{cMpc}}}

\newcommand{\dperp}{\ensuremath{\langle d_{\perp} \rangle}}

\newcommand{\ang}{\ensuremath{\mathrm{\AA}}}

\newcommand{\waveion}[3]{\ion{#1}{#2} $\lambda$#3}

\newcommand{\zbg}{\ensuremath{z_\mathrm{bg}}}
\newcommand{\za}{\ensuremath{z_\alpha}}

\newcommand{\cmd}{\mathbf{C}_\mathrm{MD}}
\newcommand{\cdd}{\mathbf{C}_\mathrm{DD}}

\newcommand{\deltarec}{\ensuremath{\delta_F^{\mathrm{rec}}}}

\shorttitle{CLAMATO IGM Tomography DR1}
\shortauthors{Lee et. al.}

\begin{document}

\title{First Data Release of the COSMOS Lyman-Alpha Mapping And Tomography Observations:\\ 
3D Lyman-$\alpha$ Forest Tomography
at $2.05 < \lowercase{z} < 2.55$}
\author{Khee-Gan Lee}
\altaffiliation{Hubble Fellow}
\affiliation{Lawrence Berkeley National Laboratory, 1 Cyclotron Road, Berkeley, CA 94720, USA}

\author{Alex Krolewski}
\affiliation{Department of Astronomy, University of California at Berkeley, New Campbell Hall,
Berkeley, CA 94720, USA}

\author{Martin White}
\affiliation{Department of Astronomy, University of California at Berkeley, New Campbell Hall,
Berkeley, CA 94720, USA}
\affiliation{Lawrence Berkeley National Laboratory, 1 Cyclotron Road, Berkeley, CA 94720, USA}

\author{David Schlegel}
\affiliation{Lawrence Berkeley National Laboratory, 1 Cyclotron Road, Berkeley, CA 94720, USA}

\author{Peter E.\ Nugent}
\affiliation{Lawrence Berkeley National Laboratory, 1 Cyclotron Road, Berkeley, CA 94720, USA}
\affiliation{Department of Astronomy, University of California at Berkeley, New Campbell Hall,
Berkeley, CA 94720, USA}

\author{Joseph F.\ Hennawi}
\affiliation{Department of Physics, Broida Hall, University of California at Santa Barbara, Santa Barbara, CA 93106, USA}

\author{Thomas M\"uller}
\affiliation{Max Planck Institute for Astronomy, K\"{o}nigstuhl 17, D-69117 Heidelberg, Germany}

\author{Richard Pan}
\affiliation{Department of Astronomy, University of California at Berkeley, New Campbell Hall,
Berkeley, CA 94720, USA}

\author{J.\ Xavier Prochaska}
\affiliation{Department of Astronomy and Astrophysics, University of California at Santa Cruz, 1156 High Street, Santa Cruz, CA 95064, USA}
\affiliation{University of California Observatories, Lick Observatory, 1156 High Street, Santa Cruz, CA 95064, USA}

\author{Andreu Font-Ribera}
\affiliation{Department of Physics and Astronomy, University College London, Gower Street, London, WC1E 6BT, UK}

\author{Nao Suzuki}
\affiliation{Kavli Institute for the Physics and Mathematics of the Universe (IPMU), The University of Tokyo, 
Kashiwano-ha 5-1-5, Kashiwa-shi, Chiba, Japan}

\author{Karl Glazebrook}
\affiliation{Swinburne University of Technology, Victoria 3122, Australia}

\author{Glenn G.\ Kacprzak}
\affiliation{Swinburne University of Technology, Victoria 3122, Australia}

\author{Jeyhan S.\ Kartaltepe}
\affiliation{School of Physics and Astronomy, Rochester Institute of Technology, 84 Lomb Memorial Drive, Rochester, NY 14623, USA }

\author{Anton M.\ Koekemoer}
\affiliation{Space Telescope Science Institute, 3700 San Martin Drive, Baltimore MD 21218, USA}

\author{Olivier Le F\`evre}
\affiliation{Aix Marseille Universit\'e, CNRS, LAM (Laboratoire d'Astrophysique  de Marseille) UMR 7326, 13388, Marseille, France}

\author{Brian C.\ Lemaux}
\affiliation{Department of Physics, University of California, Davis, One Shields Ave., Davis, CA 95616, USA}
\affiliation{Aix Marseille Universit\'e, CNRS, LAM (Laboratoire d'Astrophysique  de Marseille) UMR 7326, 13388, Marseille, France}

\author{Christian Maier}
\affiliation{University of Vienna, Department of Astrophysics, Tuerkenschanzstrasse 17, 1180 Vienna, Austria}

\author{Themiya Nanayakkara}
\affiliation{Leiden Observatory, Leiden University, P.O. Box 9513, 2300 RA Leiden, The Netherlands}

\author{R.\ Michael Rich}
\affiliation{Department of Physics and Astronomy, University of California at Los Angeles, Los Angeles, CA 90095, USA}

\author{D.\ B.\ Sanders}
\affiliation{Institute for Astronomy,
University of Hawaii,
2680 Wooodlawn Drive,
Honolulu, HI  96822,  USA}

\author{Mara Salvato}
\affiliation{Max Planck Institute for Extraterrestrial Physics, Gie{\ss}enbachstra§e 1, 85741 Garching bei M\"unchen, Germany}

\author{Lidia Tasca}
\affiliation{Aix Marseille Universit\'e, CNRS, LAM (Laboratoire d'Astrophysique  de Marseille) UMR 7326, 13388, Marseille, France}

\author{Kim-Vy H. Tran}
\affiliation{School of Physics, University of New South Wales, Kensington,
  Australia}
\affiliation{George P. and Cynthia Woods Mitchell Institute for Fundamental Physics and Astronomy, and Department of Physics and Astronomy, Texas A\&M University, College Station, TX 77843, USA}

\correspondingauthor{Khee-Gan Lee}
\email{kglee@lbl.gov}

\begin{abstract}
Faint star-forming galaxies at $z\sim 2-3$ can be used as alternative
background sources to probe the Lyman-$\alpha$ forest in addition to quasars, yielding high
sightline densities that enable 3D tomographic reconstruction of the foreground absorption field. 
Here, we present the first data release from the COSMOS Lyman-Alpha Mapping And Tomography Observations
(CLAMATO) Survey, which was conducted with the LRIS spectrograph on the Keck-I telescope. Over an observational footprint of 0.157$\mathrm{deg}^2$ within the COSMOS field, we used 240 galaxies and quasars at $2.17<z<3.00$, with a mean comoving transverse separation of $2.37\,\hMpc$, as background sources probing
the foreground Lyman-$\alpha$ forest absorption at $2.05<z<2.55$. The Lyman-$\alpha$ forest
data was then used to create a Wiener-filtered tomographic reconstruction over a comoving volume of $3.15\,\times 10^5\,h^{-3}\,\mathrm{Mpc^3}$ with an effective smoothing scale of $2.5\,h^{-1}\,\mathrm{Mpc}$. 
In addition to traditional figures, this map is also presented as a virtual-reality visualization 
and manipulable interactive figure.
We see large overdensities and underdensities that visually agree with the distribution of coeval galaxies from spectroscopic redshift surveys in the same field, including overdensities associated with several recently-discovered galaxy protoclusters in the volume. 
{Quantitatively, the map signal-to-noise is $\mathrm{S/N^{wiener}} \approx 3.4$ over a $3\,h^{-1}\mathrm{Mpc}$ top-hat
kernel based on the variances estimated from the Wiener filter.} This data release includes the redshift catalog, reduced spectra, extracted Lyman-$\alpha$ forest
pixel data, and reconstructed tomographic map of the absorption. These can be downloaded from Zenodo (\url{https://doi.org/10.5281/zenodo.1292459}).
\end{abstract}

\section{Introduction}
The Lyman-$\alpha$ (Ly$\alpha$) forest absorption from residual, diffuse,
\ion{H}{1} in the intergalactic medium (IGM) is a well-established tracer of
cosmological large-scale structure \citep[e.g.,][]{croft:1998,mcdonald:2006, slosar:2011, 
busca:2013}.
In particular, since the hydrogen Ly$\alpha$ transition (restframe wavelength $\lambda=1215.67\,\ang$) 
redshifts into
the optical atmospheric window at $z\gtrsim 2$, this makes the \lyaf{} 
a particularly important probe at redshifts that are otherwise challenging
to access through methods such as galaxy redshift surveys or 
gravitational weak lensing, which at time of writing are typically limited to $z<1$.

As the brightest ultraviolet sources in the distant universe, 
quasars have been the traditional background objects against which
the absorption of the IGM \lyaf{} can be studied along the 
foreground lines-of-sight. Due to the comparative rarity of quasars on the sky, 
however, these studies have generally been confined to one-dimensional
lines-of-sight directly in front of each quasar \citep[but see][ for early 
studies using closely-separated quasar sightlines]{dodorico:2006, rollinde:2003}.

More recently, the \lyaf{} component of the BOSS survey \citep{eisenstein:2011, dawson:2013} 
has systematically pursued
sufficiently high number densities of $z>2$ quasars such that
it becomes possible to cross-correlate the absorption seen 
in different quasar sightlines \citep{slosar:2011}, although the mean transverse
separation between sightlines is relatively large 
($\dperp \sim 20\,\hMpc$). This was, however, more than sufficient
for achieving BOSS's primary survey goal of measuring the 
the baryon acoustic oscillation signal in the
3-dimensional (3D) \lyaf{} clustering \citep{busca:2013, slosar:2013, kirkby:2013,
 font-ribera:2014a,delubac:2015, bautista:2017, du-mas-des-bourboux:2017}.

By targeting fainter background sources than the $g<22$ quasars
observed by BOSS, the mean sightline separation can be decreased 
to probe smaller scales, although the quasar luminosity
function is too shallow to be worth the steep increase in
observational resources needed: based on the \citet{palanque-delabrouille:2013} luminosity function, 
for example, $g < 24$ quasars at $2.4<z<2.8$ that 
can probe the $z\sim 2.3$ \lyaf{} only achieves target densities of $\sim 80\,\mathrm{deg}^2$ or mean separations of 
%
%
$\dperp \sim 7.5\,\hMpc$. In addition to quasars, 
it is possible to dramatically increase sightline densities by
targeting UV-emitting 
star-forming galaxies at $z>2$, often referred to as `Lyman-Break Galaxies' (LBGs) due to their original
selection method \citep{steidel:1996}. \citet{lee:2014} calculated, for example, that
a $g=24.5$ survey limit leads to $\sim 1500\,\mathrm{deg}^{-2}$ of
sightlines with a mean spacing of $\dperp \sim 2.5\,\hMpc$.

With background sources separated by only several transverse Mpc, it becomes 
interesting to carry out a tomographic reconstruction to recover
the 3D \lyaf{} absorption field on spatial resolutions that resolve the cosmic web. This concept was first proposed in 
\citet{pichon:2001} and \citet{caucci:2008}, while \citet{lee:2014} studied the observational feasibility and argued that present-day instrumentation should be capable of implementing IGM tomography down to scales of 2-3$\,\hMpc$ {(but see \citealt{ozbek:2016}
for an application of IGM tomography on larger-scale BOSS data)}.
Subsequently, pilot observations on the Keck telescope were reported in \citet{lee:2014a} and expanded, with additional
data, into an analysis of a $z=2.45$ galaxy protocluster that was previously discovered within the tomography field \citep{lee:2016}. 
Meanwhile,
\citet{stark:2015} and \citet{stark:2015a} used numerical simulations to quantify the utility of such IGM maps
for identifying galaxy protoclusters and cosmic voids, respectively, at $z\sim 2.5$ \citep[although see][for complementary
studies]{cai:2016,cai:2017}. 
\citet{schmittfull:2016} then showed that IGM tomographic maps could be used to refine photometric redshifts
of foreground galaxies with large halo masses.
Later, \citet{lee:2016a} demonstrated that upcoming IGM tomography surveys and facilities 
will be capable of recovering the 
geometric cosmic environments of large-scale structure (i.e.\ voids, sheets, filaments, and nodes) from the $z\sim 2.5$ IGM 
at comparable fidelity to $z\sim 0.4$ galaxy redshift survey maps. \citet{krolewski:2017} expanded this to demonstrate
that large-scale structure filaments can be sufficiently resolved by upcoming IGM tomography surveys
to allow constraints on galaxy-filament alignments with samples of $> 1000$ coeval
galaxies.

In this Supplement, we present the first public data release of the COSMOS Lyman-Alpha Mapping And 
Tomographic Observations (CLAMATO) survey\footnote{Website: \url{http://clamato.lbl.gov}}.
This is an observational program, conducted with the LRIS spectrograph \citep{oke:1995,steidel:2004} on the Keck-I telescope
designed as the first systematic attempt to observe relatively faint star-forming galaxies at $z\sim 2-3$ at high area densities 
($\sim1000\,\mathrm{deg}^{-2}$) in order to carry out \lyaf{} tomograpy of the foreground IGM. 
The current release incorporates observations over 0.157 square degrees of the COSMOS field
 obtained with the Keck-I telescope from 2014 through 2017.  
 
 The primary product in this release is the tomographically
  reconstructed 3D map of the $2.05<z<2.55$  \lyaf{} absorption derived from 240 background galaxies and QSOs 
  within the field, 
  but we also include the spectra and estimated redshifts of 437 objects that were successfully reduced.
  The various products have been made available in a public webpage\footnote{\url{https://doi.org/10.5281/zenodo.1292459}}, 
  and is described in Appendix~\ref{app:dr}.
  
 This paper will act as a reference for multiple science analyses with the CLAMATO data that are currently in preparation, 
 including the first detection of
  cosmic voids at $z>2$ \citep{krolewski:2017a}, the cross-correlation of \lyaf{} absorption with foreground galaxies from various spectroscopic redshift
  catalogs in the same field, and the analysis of the multiple clusters and protoclusters that 
  fall within our current volume.  {While the current footprint is likely too small for cosmic web
  analyses due to boundary effects \citep{lee:2016} , it should be sufficient to begin the first attempts on studying the scalar properties
  of large-scale structure at this epoch.}  This data set is also intended as a value-added resource for other researchers studying this 
  heavily-observed cosmic volume, as well as a reference data set to prepare   {the Ly$\alpha$ forest tomography science cases
for upcoming instruments such as the Subaru Prime Focus Spectrograph \citep{sugai:2015}, 
  Maunakea Spectroscopic Explorer \citep{mcconnachie:2016},
  Thirty Meter Telescope Wide-Field Optical Spectrograph \citep{skidmore:2015}, and the European Extremely Large
  Telescope Multi-Object Spectrograph \citep{hammer:2016}}.
  
  In this paper, we assume a concordance flat $\Lambda$CDM cosmology, with $\Omega_M=0.31$, 
$\Omega_\Lambda=0.69$ and $H_0 = 70\,\kms\,\mathrm{Mpc}^{-1}$. 
The exact choice of cosmology does not significantly affect our resulting tomographic reconstruction --- {
which is intended for galaxy evolution purposes} ---
since it only affects the conversion of redshift and angular separation into comoving distances.
{Future cosmological analyses would need to be more careful about the choice of cosmology, or indeed 
carry out analyses directly on the pixel data rather than using a tomographic reconstruction}.

\section{Survey Design and Target Selection}

\subsection{Survey Field}

\begin{figure*}[ht]\centering
\includegraphics[width=0.65\textwidth,clip=true, trim=20 30 20 20]{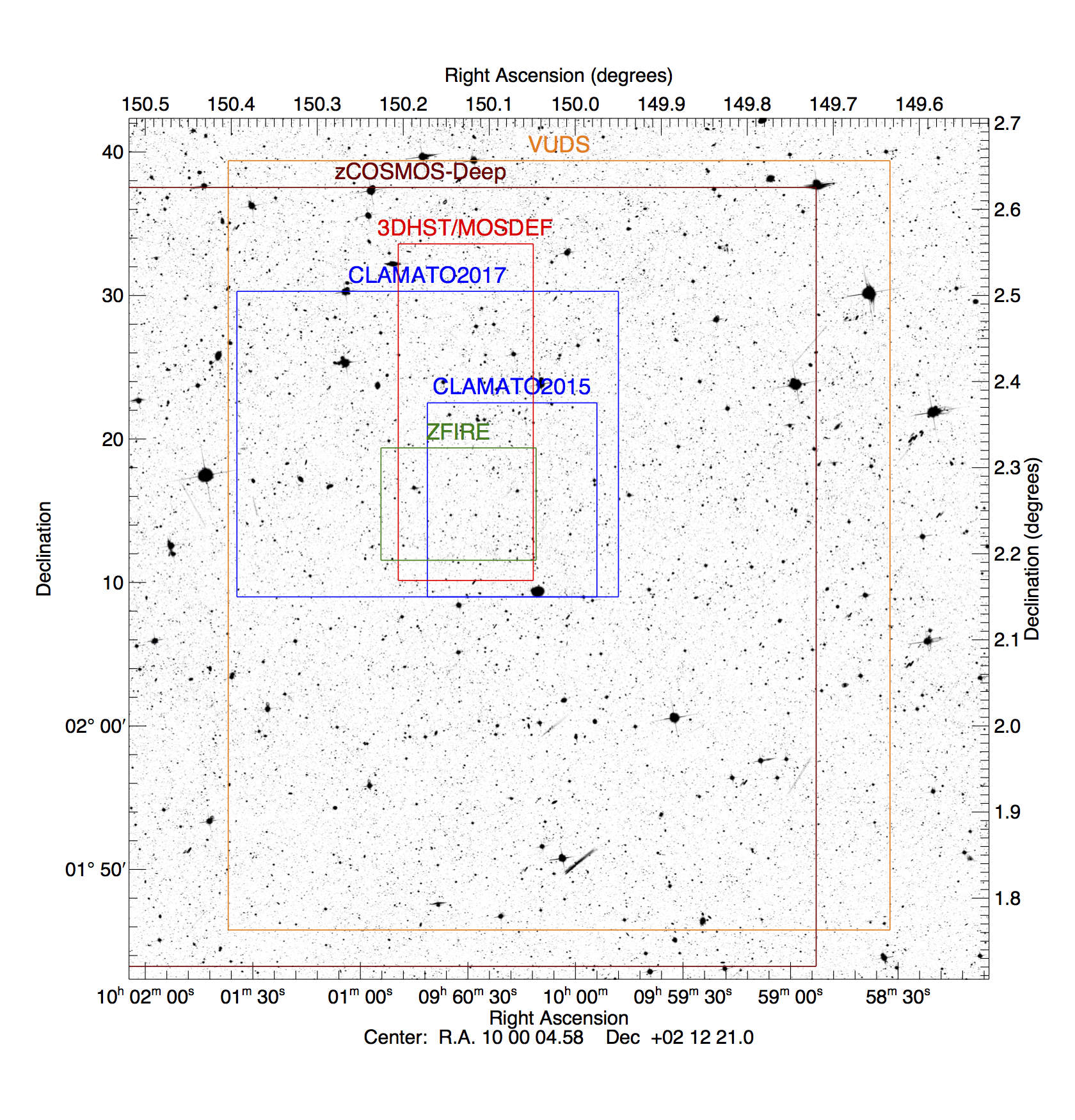}
\caption{\label{fig:survey_fields}
CLAMATO in context: this shows a \textit{Hubble Space Telescope} ACS F814W mosaic \citep{koekemoer:2007} of the central regions in the COSMOS field, 
with the footprint of the CLAMATO tomographic map indicated in blue 
(both the current paper and 2015 version, \citealt{lee:2016}). Also shown are the approximate footprints for other
spectroscopic redshift surveys that probe similar redshifts, such as 3D-HST \citep{momcheva:2016} and 
MOSDEF \citep{kriek:2015} in red, zCOSMOS-Deep \citep{lilly:2007} in brown, VUDS \citep{le-fevre:2015} in orange, and ZFIRE \citep{nanayakkara:2016} in green. {The overall ACS footprint used for the \citet{capak:2007} base $i$-band catalog is larger than the field
shown here.}
} 
\end{figure*}

As CLAMATO is the first attempt at mapping large-scale structure using IGM tomography at $z\sim 2$, 
we had to choose a well-studied
 extra-galactic field that offers sufficiently deep imaging, and ideally, spectroscopy 
 to select UV-bright star-forming galaxies with sufficient depth ($g>24$)
 as to have mean separations of $\sim 2-3'$. At the same time, we desired a large enough footprint to cover large-scale structure on 
 $\gtrsim 10\,\cMpc$ scales in the transverse dimension, i.e. an extragalactic 
 field spanning $>10'$. This left the 2$\mathrm{deg}^2$ COSMOS field \citep{scoville:2007} as the obvious candidate
 accessible from the Northern Hemisphere, which also had the additional advantage of multiple 
 deep spectroscopic surveys that cover our target redshifts, e.g. zCOSMOS \citep{lilly:2007}, VUDS \citep{le-fevre:2015}, 
 MOSDEF \citep{kriek:2015}, and ZFIRE \citep{nanayakkara:2016}. The location of these fields relative to CLAMATO
 is indicated in Figure~\ref{fig:survey_fields}.  Currently, CLAMATO has fully covered the ZFIRE footprint and approximately 80\% of the MOSDEF footprint within COSMOS.
 
 \begin{figure*}[ht]\centering
\includegraphics[width=0.67\textwidth,clip=true, trim=20 10 30 30]{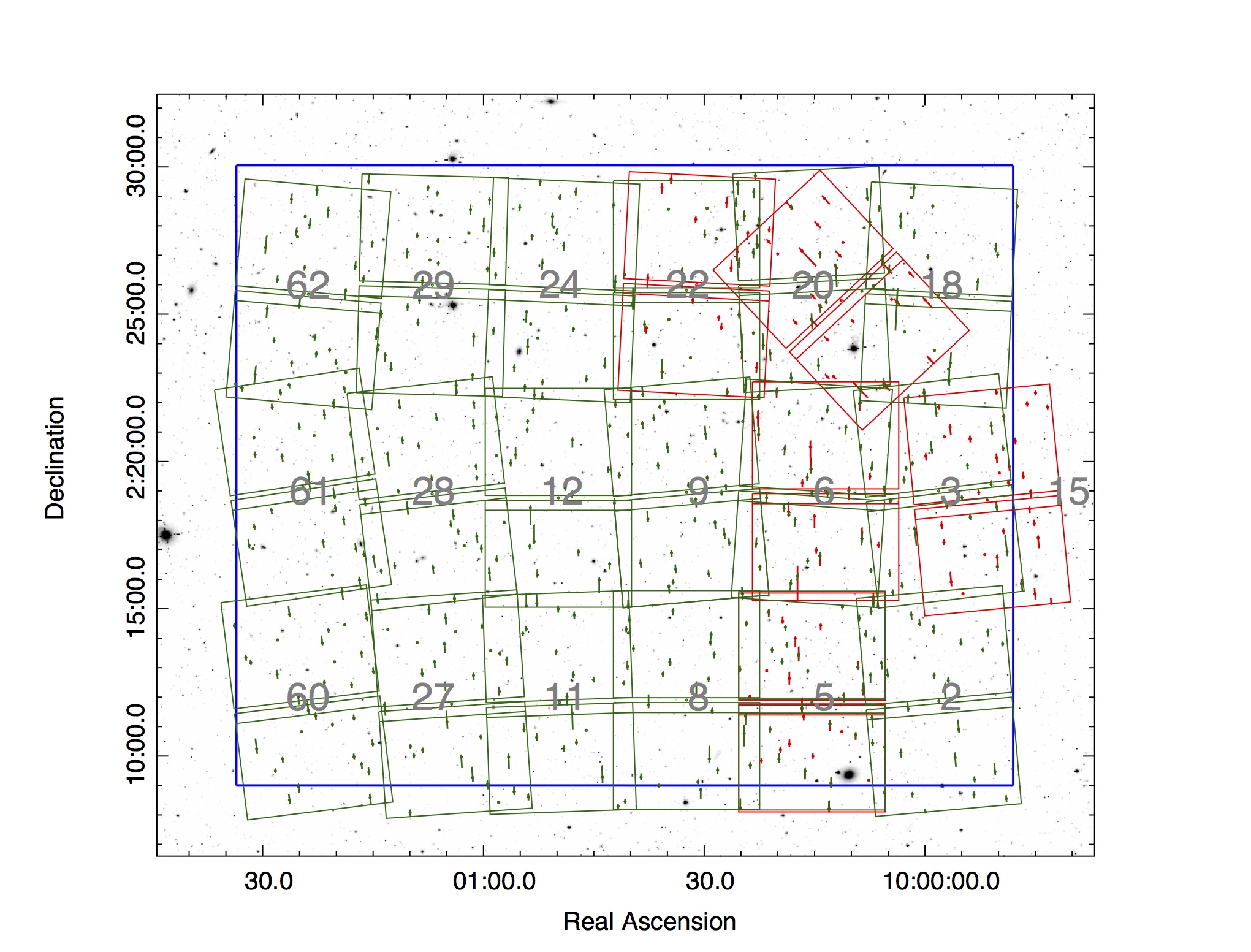}
\caption{\label{fig:slitmasks} Slits and footprints of the 23 Keck-I/LRIS slitmasks observed during the 
2014-2017 CLAMATO campaign in the 
COSMOS field, overlaid on top of the deep \textit{Hubble Space Telescope} ACS F814W mosaic of the same field 
\citep{koekemoer:2007}. The blue box indicates the footprint of the reconstructed tomographic map from the $2.15<z<2.55$
\lyaf{} absorption. Most of the slitmasks were designed to achieve a uniform survey layer (dark green), 
while several were `special'
slitmasks (red) designed to obtain additional sightlines in specific regions; see Table~\ref{tab:slitmasks}. 
The numbers in grey approximately label the field positions.
}
\end{figure*}

\subsection{Target Catalogs}
The target selection for CLAMATO is aimed at
exploiting the rich availabilty of spectroscopic and multi-wavelength imaging
data within the COSMOS field \citep{scoville:2007} in order to maximize the area density
and spatial homogeneity of $g$-band
(restframe UV at $z\sim 2-3$) sources 
that can probe the foreground \lyaf{} absorption within a narrow redshift range of $z\sim 2-3$.
The COSMOS field has high-quality multi-wavelength photometric redshifts  
\citep{ilbert:2009, laigle:2016}, as well as large numbers of spectroscopic redshifts that have already 
been obtained within 
our desired footprint and redshift range. 
We will also retarget objects that have been observed by the zCOSMOS-Deep \citep{lilly:2007} and 
VUDS \citep{le-fevre:2015} even though their spectra, in principle, cover our desired wavelength 
range ($3700\,\ang < \lambda < 4300\,\ang$). This is because the spectra from both these surveys have
 a spectral resolution of $R\sim 200$ at these wavelengths, which means that the resolution element is equivalent
to $16\,\hMpc$ line-of-sight comoving distance at $z=2.3$; this is far too coarse for our desired 
spatial resolution of several Mpc.

This data described in this paper represent three distinct 
target selection iterations:
Pilot observations (2014-2015), 2016, and 2017.
The overall target selection algorithm 
was the same over the different observing seasons, but the input catalog was
updated at the beginning of each of the aforementioned epochs
 to exploit the best-available data at that point.

Initially, we created a master raw catalog that includes a superset of objects in the 
COSMOS field with $g<25.2$ at $2.0<z<3.5$, 
which would act as a basis for target selection.
As a starting point, we use the compilation of 
available spectroscopic redshifts within the $2\,\mathrm{deg}^2$ COSMOS field by Salvato et al. (in prep),
which includes 68116 unique redshifts from all sources\footnote{We used the Apr 2015 iteration of this catalog.}. At our
 redshift of interest ($z\sim 2-3$), most spectroscopic sources
 within this compilation are from the zCOSMOS-Deep survey \citep{lilly:2007}. We then supplemented 
 this with preliminary versions of the VUDS \citep{le-fevre:2015}, MOSDEF \citep{kriek:2015}, and 
 ZFIRE \citep{nanayakkara:2016} spectroscopic 
 catalogs, as well as the 3D-HST grism redshifts \citep{momcheva:2016}.

In addition to spectroscopic redshift catalogs, we also use
 the \citet{ilbert:2009} $i$-band selected photometric 
redshift catalog, 
which in turn is based on the \citet{capak:2007} imaging multi-wavelength catalog 
in the $2\,\mathrm{deg}^2$ COSMOS field. The photometric redshifts from \citet{ilbert:2009}
exploit a wide array of multi-wavelength data with up to 30 bands ranging from the ultraviolet to radio wavelengths. 
This yields a relatively accurate redshift
estimate and low catastrophic failure rate. In 2017, we supplemented this with the \citet{davidzon:2017} 
photometric redshift catalog, which is a high-redshift optimization of the NIR-selected catalog 
of \citet{laigle:2016} and provides more
accurate photometric redshifts than \citet{ilbert:2009}. However, since this is a NIR-selected catalog {based on $z^{++}$-band
and 
$Y$$J$$H$$K_s$-band selection}, it does not provide good
completeness for restframe UV-bright objects that require an optical detection. 
We therefore continue to use the \citet{ilbert:2009} catalog to provide a baseline of objects
and simply replace the photometric redshift values by the \citet{davidzon:2017} version for objects that have a match. 
For part of the field, we were also able to use the ZFOURGE medium-band redshifts \citep{straatman:2016} that should provide 
superior photometric redshifts at our target redshift; these were also incorporated, where available, by overriding 
the \citet{ilbert:2009} and
\citet{davidzon:2017} redshifts.

\subsection{Selection Algorithm}

The target selection was then carried out as a two-step procedure: initial selection
and prioritization of possible targets, followed by slitmask design with slit assignments guided by the target priorities.
Note the difference between these steps: target \textit{selection} involves
identifying all objects that might possibly be used for our purposes and prioritizing them based on redshift, magnitude, and
probability of success (e.g.\ spectroscopic versus photometric redshift from surveys of varying quality); 
but not all of these will
be \textit{assigned} slits due to packing constraints on each slitmask.

In the selection/prioritization step, we fed the combined spectroscopic and photometric catalog to an algorithm designed to initially 
select and prioritize background $g$-band sources to homogeneously probe a fixed \lya{} absorption redshift \za.
In our case, since we aimed to probe a finite redshift range at $z\sim 2.3$, we ran the target selection algorithm at $\za = 2.25$
and $\za = 2.45$ and collated the targets.
This algorithm first divides the field into square cells of 2.75 arc-minutes on a side, approximately our desired
sightline separation. For each cell, it selects candidate background sources at redshifts 
$(1+\za) 1216/1195 -1 < \zbg < (1+\za) 1216/1040-1$, that could probe the forest absorption at \za\ in the restframe
 $1040\,\ang<\lambda<1216\,\ang$ spectral region between the \lya{} and Ly$\beta$ transitions.
It then gives the highest priority to ``bright'' sources (defined as $g<[24.2,24.4]$ at $\za =[2.25, 2.45]$, respectively)
that have spectroscopic redshifts, while faint or photometric redshift-selected
objects are down-prioritized. Due to slit-packing constraints, the algorithm deprioritizes relatively bright sources 
if another, brighter, high-confidence target is within the same cell, while fainter or photometric redshift targets might receive 
relatively high priority in the absence of other suitable background sources within its 2.75 arc-minute cell.
To take into account the possibility that slit collisions from targets in other cells  
might clobber the highest-priority source within a given cell, the algorithm selected multiple sources per cell 
(with decreasing priority) where available.
This procedure selected targets as faint as $g=25.3$ in regions with a paucity of better sources, 
but such faint targets were assigned a commensurately low priority.

The initial selection of sources, and their priority rankings from this algorithm, were then fed into the AUTOSLIT3 
software\footnote{\url{https://www2.keck.hawaii.edu/inst/lris/autoslit_WMKO.html}} in order to manually design LRIS slitmasks.
For the slitmask design, we chose slits with $1"$ width and minimum length of $6.5"$ separated by $1"$.
The initial slit assignment was automatically carred out by AUTOSLIT3 based on the priorities assigned by the initial 
target selection algorithm, which we then refined in order to maximize homogeneity of bright sources 
and uniformity of redshift coverage
within our desired $2 \lesssim \za \lesssim 2.5$ absorption redshift range. 
This manual refinement included modifying the position angle of the slitmask (up to $\pm 6-7$ 
degrees\footnote{The noteable exception is slitmask sp18L, which was designed with a $43\degr$ position angle
 in an attempt 
cover a specific gap in the sightline coverage.}) in order to mitigate
slit collisions between high-priority targets. We also overlapped the slitmasks slightly in the R.A. direction, 
in order to ensure at least $\lambda > 3700\,\ang$ spectra coverage for all sources.
For each $7' \times 5'$ LRIS slitmask, we were able to assign $\sim 20-25$ science slits. Due to slit-packing constraints
and the necessity of having at least 4 alignment stars within each slitmask, 
this in fact included only $\sim 80\%$ of high-priority targets we would have liked to observe within 
our desired redshift range --- we frequently had slit collisions between high-priority sources (or with box stars), 
while available slits
elsewhere had no high-priority targets and were assigned to low-priority targets. A higher slit-packing density
would have allowed a slight improvement in sightline density at the same depth, or an increase in the absorption
redshift range beyond the $2.05<z<2.55$ charted in this survey.

We designed a uniform set of slitmasks to cover our entire survey footprint (Figure~\ref{fig:slitmasks}), but also supplemented these 
with additional slitmasks (Table~\ref{tab:slitmasks}) --- designed and observed in subsequent observing seasons after the initial pass--- 
to increase sightline sampling
in particular regions of interest, or to make up for shortfalls in sightline density after the initial round of observations.


\section{Observations \& Data Reduction}

 \begin{deluxetable*}{l c c c c l}[htb!]
\tablecolumns{6}
\tablecaption{\label{tab:slitmasks} CLAMATO Data Release 1 Slitmasks}
\tablehead{
 Maskname\tablenotemark{a} & $\alpha$ (J2000)\tablenotemark{b} & 
 $\delta$ (J2000)\tablenotemark{b} & Exposure Time (s)
& Year Observed & Remarks }
\startdata
cpilot09 &  10 00 33.067 &  +02 20 50.58 &  7200 & 2014 &    Uniform Survey Mask      \\ 
cpilot08 &  10 00 32.404 &  +02 13 48.01 &  7200 &  2014 &   Uniform Survey Mask        \\ 
cpilot05 &  10 00 15.365 &  +02 13 47.01 &  7200 &  2015  &   Uniform Survey Mask       \\ 
cpilot06 &  10 00 14.834 &  +02 20 48.73 &  7200 &  2015 &  Uniform Survey Mask        \\ 
cpilot02 &  09 59 58.765 &  +02 13 45.55 &  7200 &  2015 &   Uniform Survey Mask        \\ 
cpilot03 &  09 59 59.014 &  +02 20 53.21 & 10800 &  2015 &  Uniform Survey Mask         \\ 
cpilot12 &  10 00 49.818 &  +02 20 40.01 & 16200 &   2014/2016 &  Uniform Survey Mask       \\ 
pc06 &  10 00 13.503 &  +02 20 53.43 &  7200 &    2015 & Targeted at $z=2.10$ Protocluster      \\ 
npc05 &  10 00 15.358 &  +02 13 43.08 & 19800 &  2016 &  Targeted at $z=2.30$ Galaxy Overdensity        \\ 
c16\_11 &  10 00 49.944 &  +02 13 43.01 &  7200 &  2016 &   Uniform Survey Mask        \\ 
c16\_24 &  10 00 49.014 &  +02 27 42.63 &  7200 &  2016 &    Uniform Survey Mask       \\ 
c16\_20 &  10 00 15.809 &  +02 28 04.78 &  7200 &   2016 &    Uniform Survey Mask      \\ 
c16\_22 &  10 00 32.398 &  +02 27 42.96 &  7200 &   2016 &    Uniform Survey Mask      \\ 
c16\_18 &  09 59 57.717 &  +02 27 32.81 &  9000 &   2016 &    Uniform Survey Mask      \\ 
c17\_27s &  10 01 04.866 &  +02 13 39.53 & 10200 &  2017 &    Uniform Survey Mask       \\ 
c17\_29 &  10 01 06.761 &  +02 27 52.92 &  7200 &  2017 &  Uniform Survey Mask        \\ 
c17\_28s &  10 01 07.846 &  +02 20 47.44 &  7200 & 2017 &   Uniform Survey Mask        \\ 
c17\_62 &  10 01 23.139 &  +02 27 33.91 & 12600 &  2017 &  Uniform Survey Mask        \\ 
c17\_61L &  10 01 25.656 &  +02 21 00.15 & 12600 & 2017 &  Uniform Survey Mask         \\ 
c17\_60L &  10 01 24.926 &  +02 13 42.49 &  9000 & 2017 &  Uniform Survey Mask         \\ 
pc22L &  10 00 30.622 &  +02 27 53.81 & 10800 & 2017 &  Targeted at $z\sim 2.5$ Cluster/Protocluster         \\ 
sp18 &  10 00 16.563 &  +02 26 51.88 & 11100 & 2017 &  Designed to plug sightline gap        \\ 
sp15l &  09 59 52.268 &  +02 20 35.06 &  8700 & 2017 &   Designed to plug sightline gap       \\ 
\enddata
\tablenotetext{a}{Mask name suffixes correspond roughly to field numbers shown in Figure~\ref{fig:slitmasks}.}
 \tablenotetext{b}{Slitmask pointing center.}
\end{deluxetable*}
 
The CLAMATO observations were carried out using the LRIS spectrograph \citep{oke:1995,steidel:2004} on the Keck-I
telescope at Maunakea, Hawai'i. The observations described in this papers were carried out in the spring semesters of 
2014-2017 via a total time allocation of 15.5 nights, of which 13.5 nights were allocated by the University of California Time Allocation Committee (TAC) and 2 nights
were from the Keck/Subaru exchange time given by the National Astronomical Observatory of Japan TAC.
Out of this overall allocation, we achieved approximately 60hrs of on-sky integration\footnote{On any given night, from 
Hawai'i,
there was at most 5.5hrs in which the COSMOS field could be observable below our threshold of airmass 1.5.}.

For CLAMATO, we focused on the LRIS blue channel which covers the $3700\,\ang < \lambda < 4400\,\ang$ wavelength
range corresponding to restframe \lya{} at $2.1\lesssim \za \lesssim2.6$, our redshifts of interest. 
All our observations used the 600-line grism blazed at $4000\,\ang$ in the blue channel, 
which offers spectral resolution of $R \equiv \lambda/\Delta\lambda \approx 1100$ with $1''$ slits. 
This translates to a spectral FWHM $\approx 4\,\ang$ or a line-of-sight spatial resolution of $3\,\hMpc$ at $z\sim2.3$, which 
is a good match for our desired sightline separation.  The red channel was used primarily to assist in object 
identification and redshift estimation. In the first two nights of the 2014 observations, we used the d500 dichroic to split
the red photons into the red camera with 600-line grating blazed at $7500\,\ang$, but this was deemed to have too short a 
wavelength coverage, and so in all subsequent observations we used the d560 dichroic with the 400-line grating
blazed at $8500\,\ang$. 
This allowed better spectral coverage in the red (up to $\approx 9000\,\ang$) at the expense of lower spectral
resolution, which is still sufficient for spectral identification.

The observations were carried out at a mean seeing of $\approx 0.7''$. In seeing conditions of $<0.8''$ seeing, we typically
exposed for a total of 7200s per `normal' survey slitmask,
 but this was increased up to 14400s in sub-optimal seeing in order to achieve roughly homogeneous minimum
signal-to-noise over all our data. 
For `special' slitmasks designed to plug gaps in sightline coverage from the `normal' slitmasks, we 
integrated longer to build up signal-to-noise on fainter background sources, up to 19800s (however, many of these
longer integrations were to make up for inferior seeing conditions).
Seeing conditions that were consistently above $1.0''$ was deemed unuseable for CLAMATO, at which point
we moved on to backup targets unrelated to IGM tomography. The individual exposures were typically 1800s on the blue channel 
but only 860s on the red channel
in order to reduce the number of cosmic ray hits in the latter's thick fully-depleted CCDs \citep{rockosi:2010}. 
In practice, we carried out quick reductions during the observing run to gauge data quality, 
and occasionally obtained further integrations on a slitmask if the signal-to-noise was considered inadequate after
the standard 7200s.  A number
of the objects were assigned slits in the overlap region between two (or more) slitmasks, and therefore received 
considerably more exposure time. Over this observing campaign, we observed 18 `regular' slitmasks over the
survey footprint, and also 5 `special' slitmasks (Table~\ref{tab:slitmasks} and Figure~\ref{fig:slitmasks}). 

The data were reduced with the LowRedux routines from the XIDL software 
package\footnote{\url{http://www.ucolick.org/~xavier/LowRedux}}. After the initial flat-fielding, 
slit definition and sky subtraction, we co-added the 2D images of the 
individual exposures before tracing the 1D spectra. We found that this helps the extraction of faint source spectra, 
rather than co-adding the 1D spectra extracted from the individual exposure frames.
Due to instrument flexure, this was generally feasible only with exposures observed within the same night or adjacent nights. 
In cases where data from different observing epochs could not directly be co-added in 2D, the spectra from each epoch were co-added in 1D after
extraction and flux calibration. There were 56 objects that were targeted in more than one slitmask, and their 1D spectra
were similarly co-added in the same way after initial reduction and extraction. One particular object (ID\# 00954) received as much as
11.5hrs of integration from being in the overlap region of 4 slitmasks.

\begin{figure}\centering
 \includegraphics[width=0.46\textwidth]{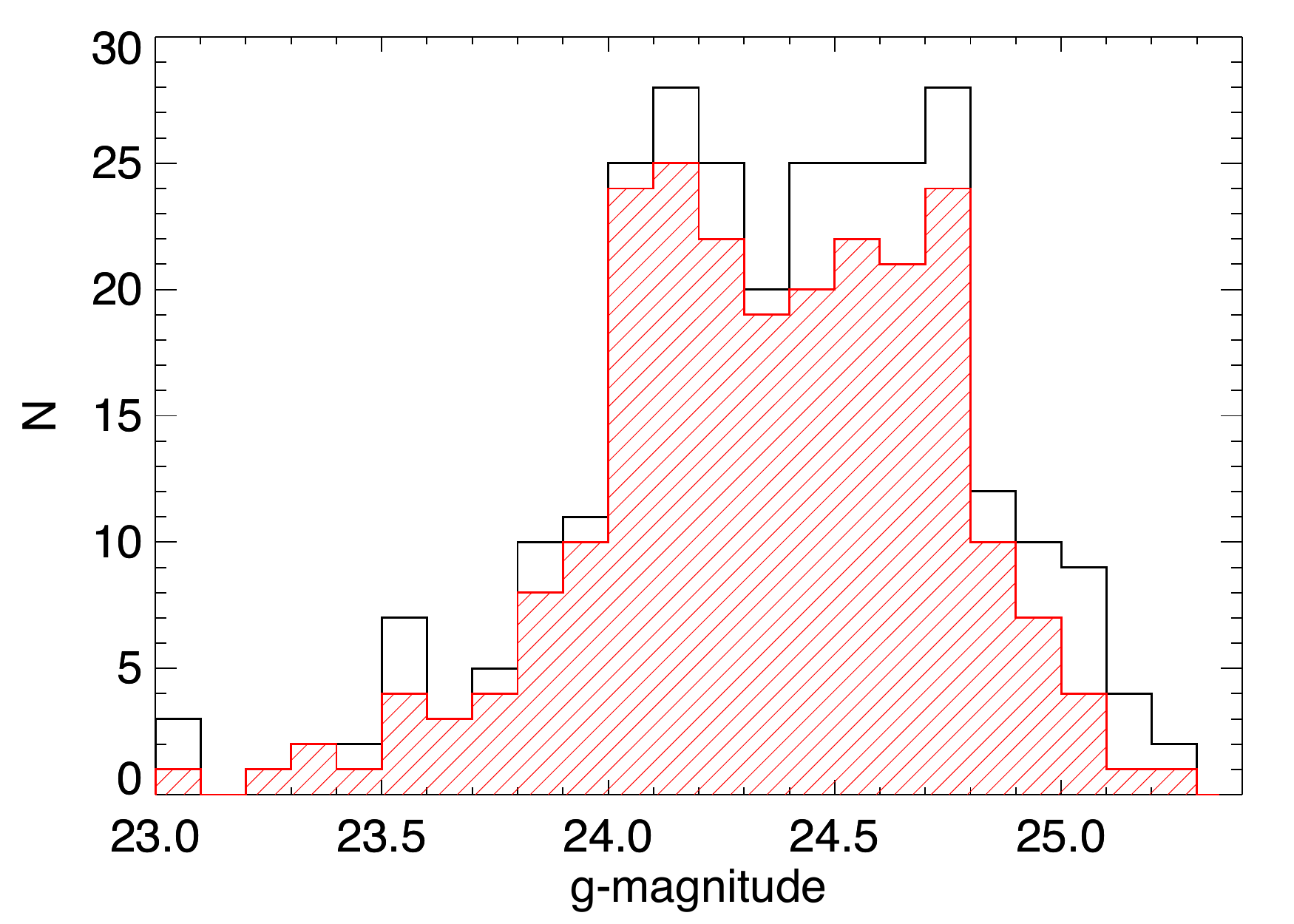}\\ 
  \includegraphics[width=0.46\textwidth]{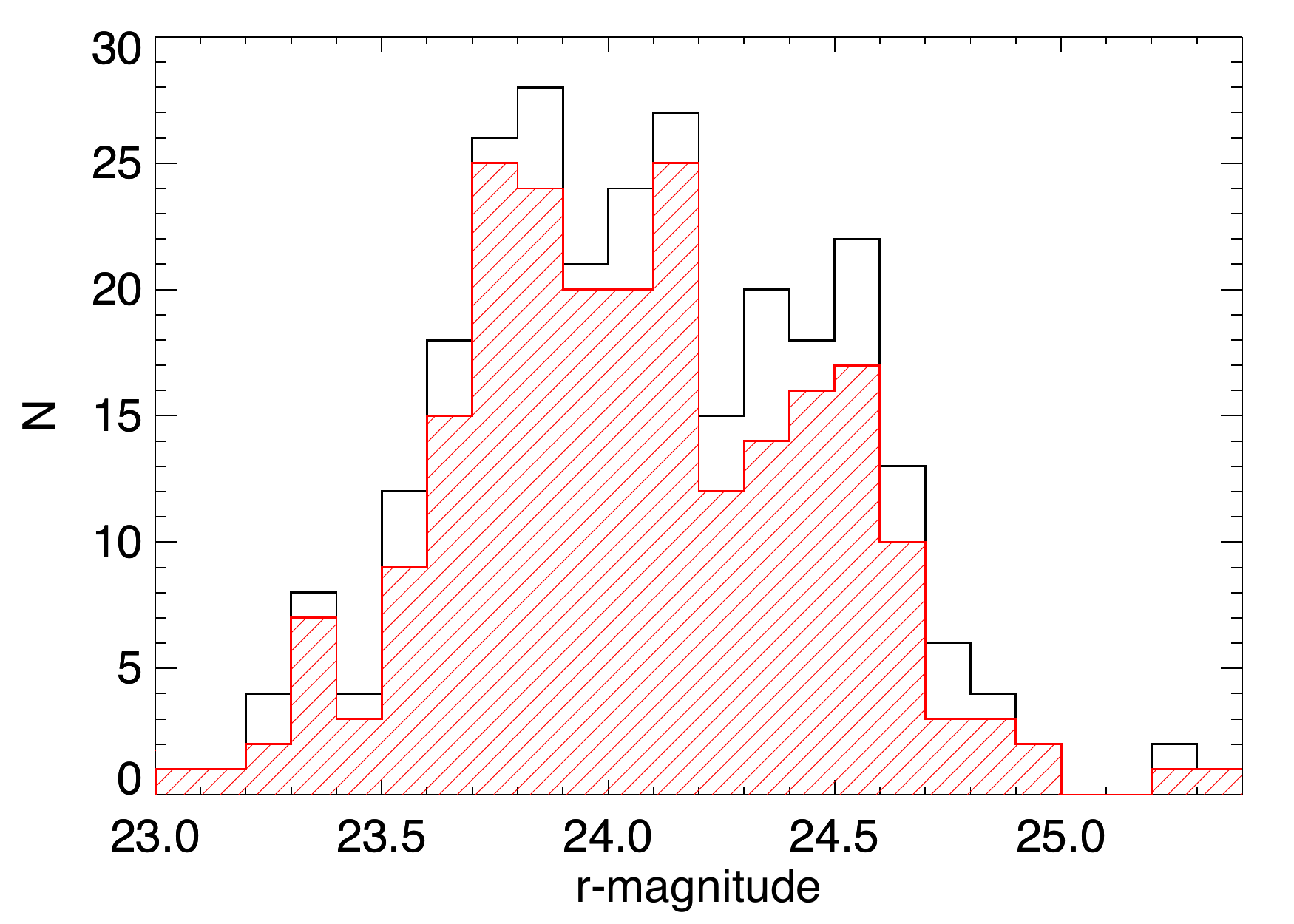}
  \caption{\label{fig:maghist}
  Magnitude distribution of CLAMATO objects with high-confidence ($>3$) redshift identifications, 
  showing $g$-magnitude at top and $r$-magnitude at bottom.
  In both cases, the red histogram indicates objects that were subsequently 
  used as background sources for the
  \lyaf{} tomographic reconstruction. Small numbers of bright ($<23$rd magnitude) sources have been
  omitted in these axes.
  }
 \end{figure}

\begin{figure}
\includegraphics[width=0.5\textwidth]{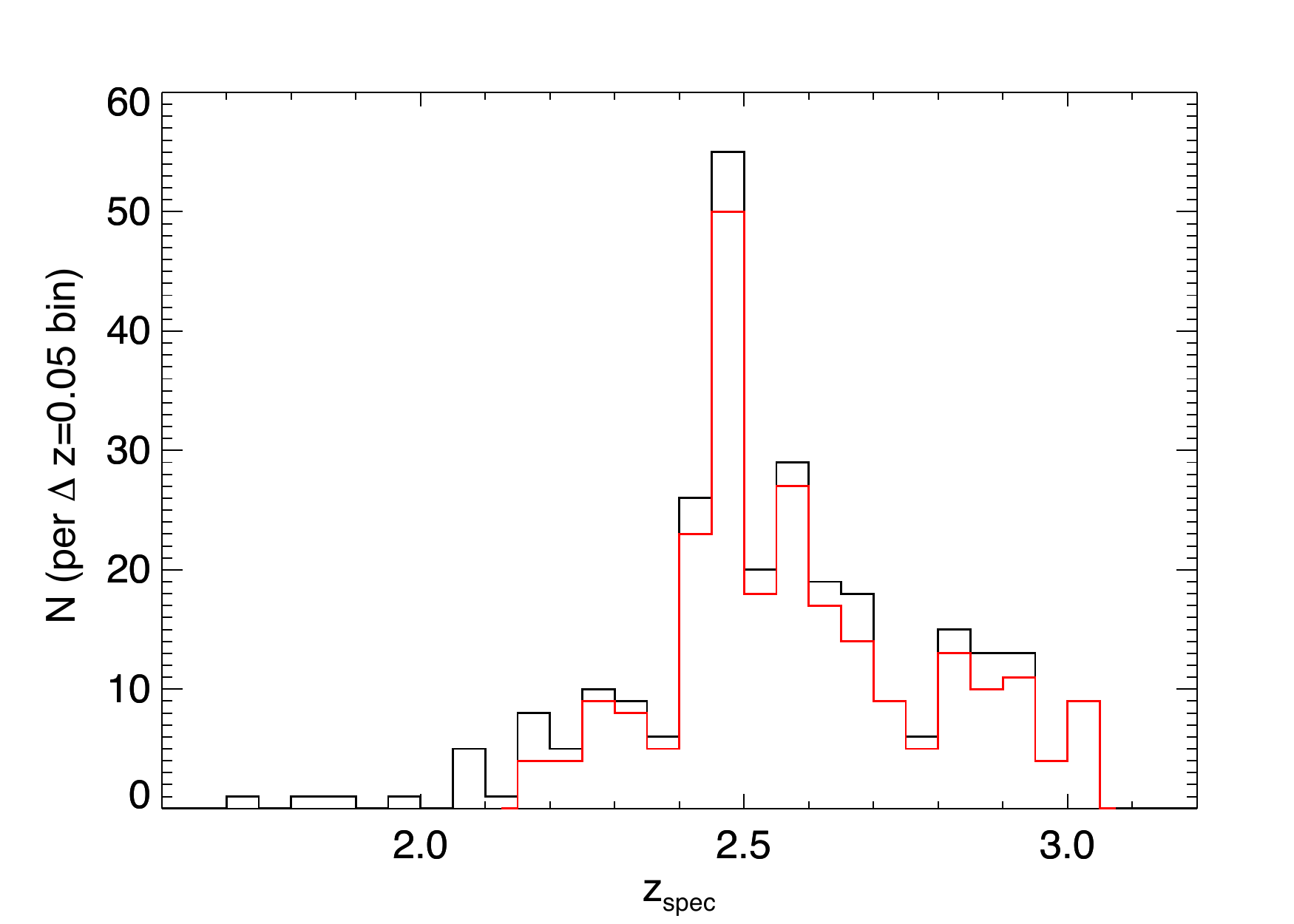} 
\caption{\label{fig:source_zhist}
Redshift distribution of well-identified ($\geq 3$ confidence rating) spectra in the current CLAMATO data release, shown
as the black histogram with redshift bins of $\Delta(z)=0.05$. 
The red histogram indicades background sources that were actually used to tomographicaly reconstruct
 the foreground \lyaf{} at $2.05<\za<2.55$.  These plot axes leave out 8 objects at $z<1.6$ and 1 object at $z>3.2$.
}
\end{figure}

\begin{figure}\centering
\includegraphics[width=0.5\textwidth]{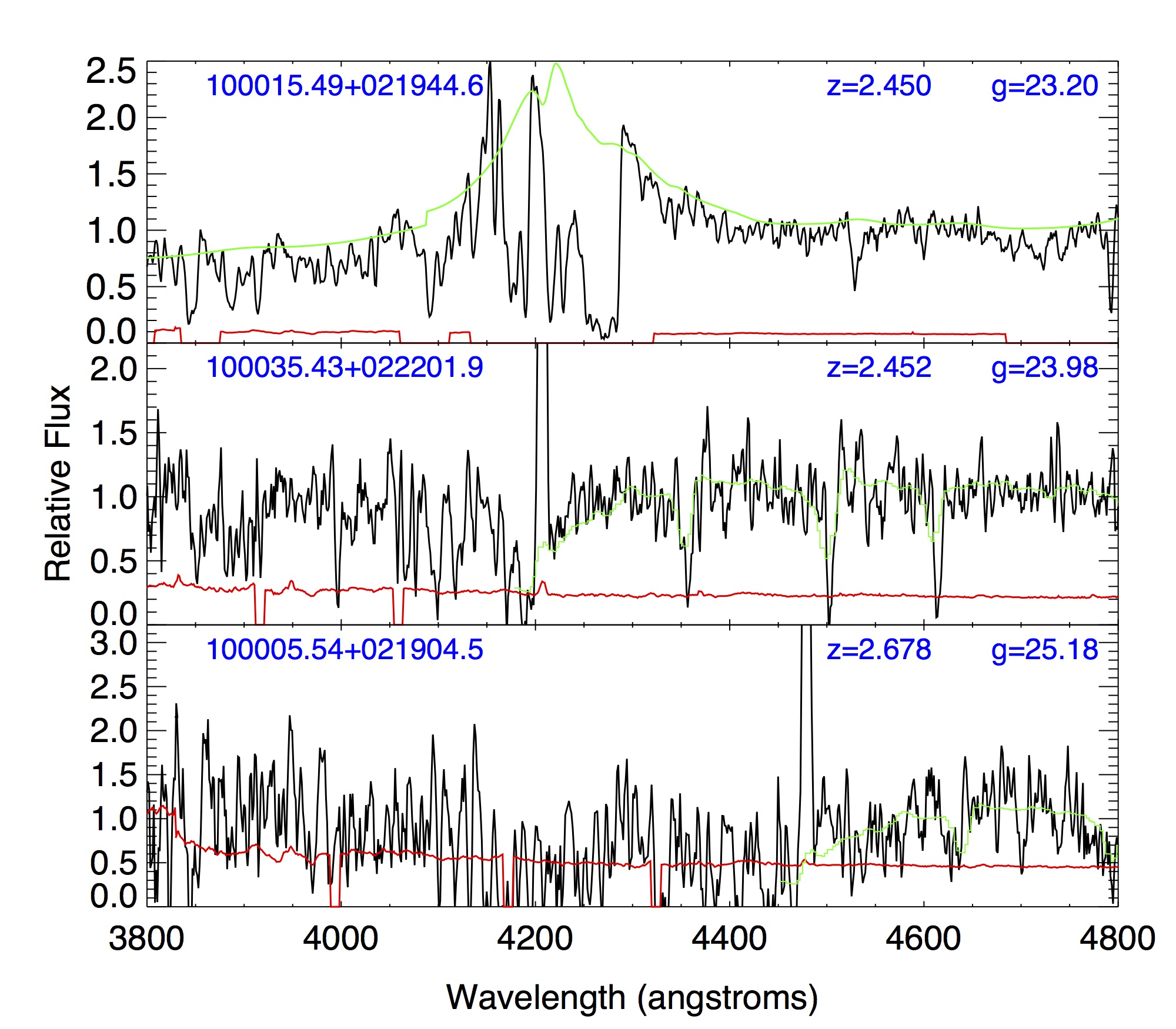}
\caption{\label{fig:spec_eg}
Examples of the reduced high-redshift spectra from our data set. The object at the top is a quasar, while the
others are LBGs with \lya{} emission. For clarity, the spectra have been smoothed with a 3-pixel tophat filter.
The galaxy at the bottom is among our faintest objects, and has marginally sufficient signal-to-noise in the \lyaf{}
to contribute to our tomographic reconstruction thanks to an above-average 6hrs of exposure over multiple slitmasks.
}
\end{figure}

From the 23 unique slitmasks observed in the 2014-2017 CLAMATO campaign (Table~\ref{tab:slitmasks}), we successfully reduced and extracted
 437 spectra from the blue channel (not including 19 spectra from unrelated `filler' programs). We also
 reduced the red channel but the extraction proved to be more challenging than in the blue, yielding only 185 corresponding red spectra. The spectra were visually inspected and compared with common line transitions and spectral templates,
 particularly the \citet{shapley:2003} composite LBG template, 
 in order to determine their identity and redshift. For each spectrum, we assigned a confidence ranking of 0-4, where 0 implies
 no attempt at an identification (usually due to corrupted spectra or little/non-existent source flux), 1 is a guess, 2 is a low-confidence redshift, 
 3 denotes a reasonable confidence, while 4 is a high-confidence redshift derived from multiple spectral features. 
 Out of the 437 reduced spectra, 289 spectra had confident identifications ($\geq$3 confidence rating) of which 277
 were at redshifts $z>2$ (Figure~\ref{fig:source_zhist}). 
 These high-redshift sources can be further classified into 262 galaxies ($95\%$) and 15 broad-line quasars (5\%). 
 Our main rationale for classifying a source as either a galaxy or quasar is to determine their continuum-fitting method;
 therefore we classified any source that showed intrinsic absorption lines at restframe 
 $\lambda > 1216\,\ang$ as a galaxy
 even if it shows a broad \lya{} emission line indicative of AGN activity. 
 Table~\ref{tab:tomo_obj} tabulates our full catalog of extracted sources, 
 while examples of the high-redshift spectra are shown in Figure~\ref{fig:spec_eg}. 
 The $g$- and $r$-magnitude (AB) distributions of high-confidence spectra are shown in Figure~\ref{fig:maghist}.
 The median magnitudes of all the high-confidence spectra, regardless of redshift, 
 are $\langle g \rangle = 24.38$ and $\langle r \rangle=24.03$,
 respectively. As we shall discuss later (\S~\ref{sec:tomo}), we will be quite aggressive in selecting
 background sources for \lyaf{} reconstruction, and therefore the median magnitudes of the final background
 sightline sample are only slightly brighter than this: $\langle g \rangle = 24.34$ and $\langle r \rangle = 24.02$.
 
 
 The relatively low rate (65\%) of confidently-identified objects relative to the extracted spectra
  is because we filled any spare slits in 
 our slitmasks with faint low-priority targets, which often resulted in spectra too noisy to be identified with confidence.
 However, of the spectra that did indeed get identified at high confidence, the yield of high-redshift ($z>2$) objects is 
 excellent (96\%), reflecting our strategy of
 retargeting spectroscopic catalogs and the high quality of the photometric redshifts of those
  that had no prior spectroscopic redshifts.

 \begin{deluxetable*}{l c c c c c c c c c c c c}[htb!]
\tablecolumns{13}
\tablecaption{\label{tab:tomo_obj} CLAMATO Data Release 1 Source Catalog}
\tablehead{
ID\# & $\alpha$ (J2000)\tablenotemark{a} & $\delta$ (J2000)\tablenotemark{a} & $g$-mag\tablenotemark{a} & 
$z_{\rm photo}$\tablenotemark{b} & $z_{\rm spec}$ 
& Conf\tablenotemark{c} & Type & $t_{\rm exp}$ (s) & Tomo\tablenotemark{d} & $\mathrm{S/N_{Ly\alpha1}}$\tablenotemark{e} &
$\mathrm{S/N_{Ly\alpha2}}$\tablenotemark{f}  & $\mathrm{S/N_{Ly\alpha3}}$\tablenotemark{g} }
\startdata
00762 &  10 01 00.905 &  +02 17 27.96 & 24.21 &   1.11 &  2.465 & 2 & GAL & 
7200
 & N & \nodata & \nodata & \nodata \\
00765 &  10 01 00.297 &  +02 17 02.58 & 24.64 &   2.93 &  2.958 & 4 & GAL & 
7200
 & Y & \nodata & \nodata &  2.5 \\
00767 &  10 01 14.934 &  +02 16 45.23 & 24.73 &   0.21 &  2.578 & 3 & GAL & 
12600
 & Y &  3.1 &  3.2 &  3.0 \\
00771 &  10 01 06.870 &  +02 16 23.38 & 24.70 &   2.58 &  2.530 & 3 & GAL & 
7200
 & Y &  1.6 &  1.9 &  2.0 \\
00780 &  10 01 14.359 &  +02 15 15.84 & 24.28 &   0.08 &  0.082 & 2 & GAL & 
7200
 & N & \nodata & \nodata & \nodata \\
00783 &  10 01 07.412 &  +02 14 58.31 & 24.27 &   2.59 &  2.579 & 4 & GAL & 
10200
 & Y &  4.1 &  4.5 &  4.7 \\
00784 &  10 01 15.952 &  +02 14 48.41 & 22.02 &   2.47 &  2.494 & 4 & QSO & 
9000
 & Y & 11.5 & 13.1 & 22.1 \\
00785 &  10 01 05.138 &  +02 14 41.21 & 24.51 &   2.44 &  2.506 & 4 & GAL & 
10200
 & Y &  2.1 &  2.5 &  2.8 \\
00787 &  10 01 21.083 &  +02 14 16.48 & 24.41 &   2.62 &  2.491 & 3 & GAL & 
9000
 & N &  0.8 &  1.0 &  1.1 \\
00788 &  10 01 33.860 &  +02 14 25.19 & 24.24 &   2.62 &  2.738 & 3 & GAL & 
9000
 & Y & \nodata &  1.6 &  1.8 \\
\enddata
 \tablenotetext{a}{Source positions and magnitudes from \citet{capak:2007}.}
 \tablenotetext{b}{Photometric redshift estimate; see text for details.}
 \tablenotetext{c}{Redshift confidence grade, similar to that described in \citet{lilly:2007} but without fractional grades.}
 \tablenotetext{d}{Usage in \lya{} forest tomographic reconstruction}
 \tablenotetext{e}{Median per-pixel spectral continuum-to-noise ratio within the $2.05<\za<2.15$ \lyaf{}.}
  \tablenotetext{f}{Median per-pixel spectral continuum-to-noise ratio within the $2.15<\za<2.35$ \lyaf{}.}
   \tablenotetext{g}{Median per-pixel spectral continuum-to-noise ratio within the $2.35<\za<2.55$ \lyaf{}.}
 \tablecomments{Table 1 is published in its entirety in the machine-readable format.
      A portion is shown here for guidance regarding its form and content.}
\end{deluxetable*}

 For $z>2$ LBGs, redshifts estimated from restframe-UV spectral features are known to have offsets from the 
 `true' systemic redshifts as determined from restframe optical nebular emission lines \citep{steidel:2010, rakic:2011}. 
 For CLAMATO, the redshift estimation of the spectra is intended to achieve two purposes: selection of the foreground 
 \lyaf{} absorption from the spectral region between the intrinsic \lya{} and Ly$\beta$ wavelengths of the background source,  
and masking of the small number of intrinsic absorption lines within the \lyaf{}. The selection of the \lyaf{} pixels is 
relatively insensitive to the precise systemic redshift {of the background source}, but the masking of the intrinsic absorption lines is carried out with
narrow spectral ranges. We therefore choose to estimate the source redshift, wherever possible, based on the restframe
$\lambda>1216\,\ang$ absorption lines since this allows the best masking of the absorption lines within the 
LBG forest.

The estimated redshifts for all 437 sources are provided in the online version of Table~\ref{tab:tomo_obj}, including low-confidence objects. We have also made all the reduced spectra available for download; see Appendix~\ref{app:dr} for details.

\section{Tomographic Reconstruction}\label{sec:tomo}

\begin{figure*}[ht]\centering
\includegraphics[width=0.72\textwidth,clip=true, trim=0 10 30 30]{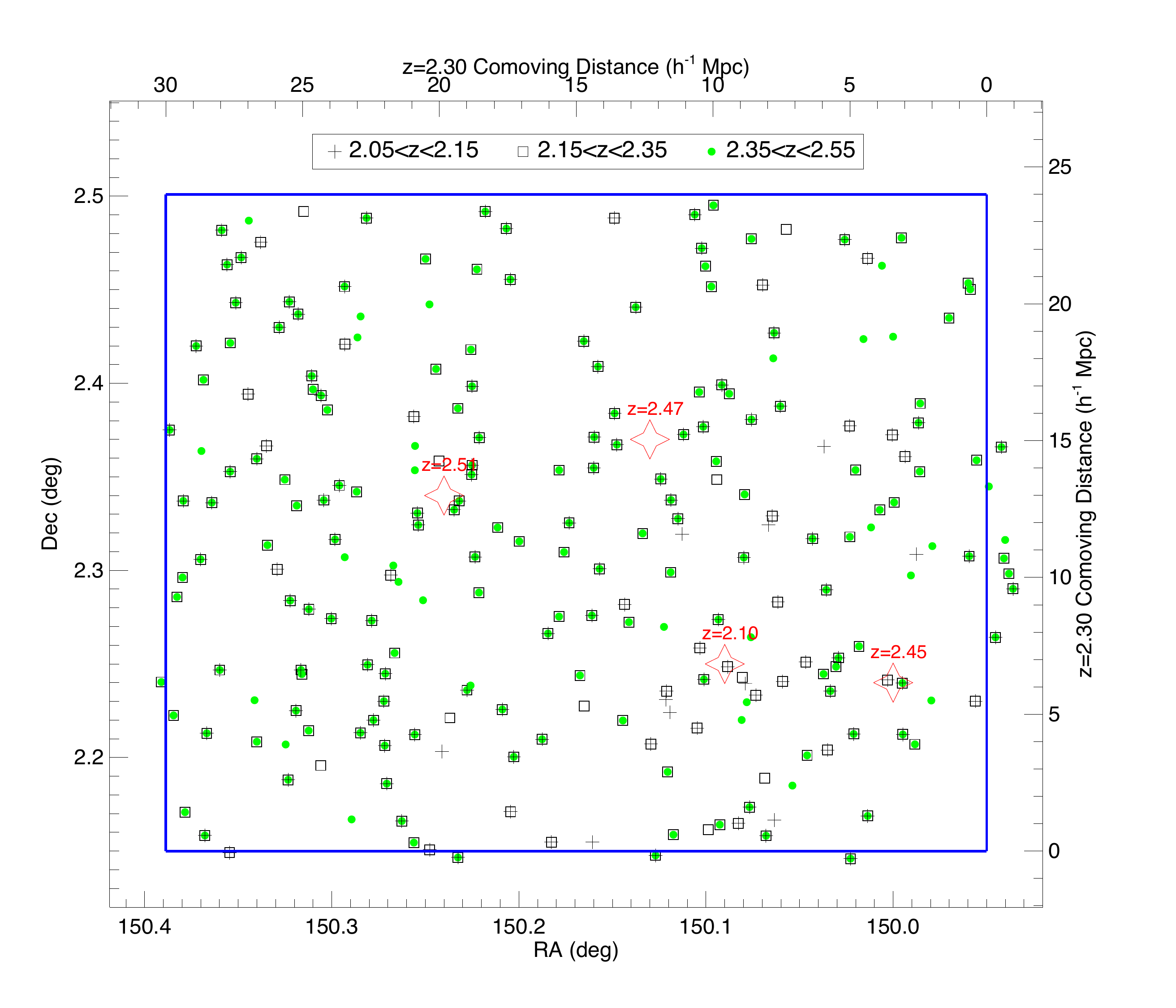}
\caption{\label{fig:sightlines_pos}
Angular position of the \lyaf{} sightlines used to tomographically reconstruct the \lyaf{} at $2.05<z<2.55$. 
The different symbols denote coverage over different redshift ranges.
Some background sources have the correct redshift to cover large ranges of our targeted foreground
redshift range
and are therefore indicated by multiple symbols.
We have also marked with red diamonds the angular position of several known overdensities, 
at $z=2.10$ \citep{spitler:2012, nanayakkara:2016}, $z=2.44$ \citep{diener:2015, chiang:2015}, 
$z=2.47$ \citep{casey:2015}, and $z=2.51$ \citep{wang:2016}.  The top and right-hand axes
denote the transverse comoving distances in the coordinates of our tomographic map grid.
} 
\end{figure*}

\begin{figure}
\includegraphics[width=0.48\textwidth,clip=true,trim=10 0 0 0]{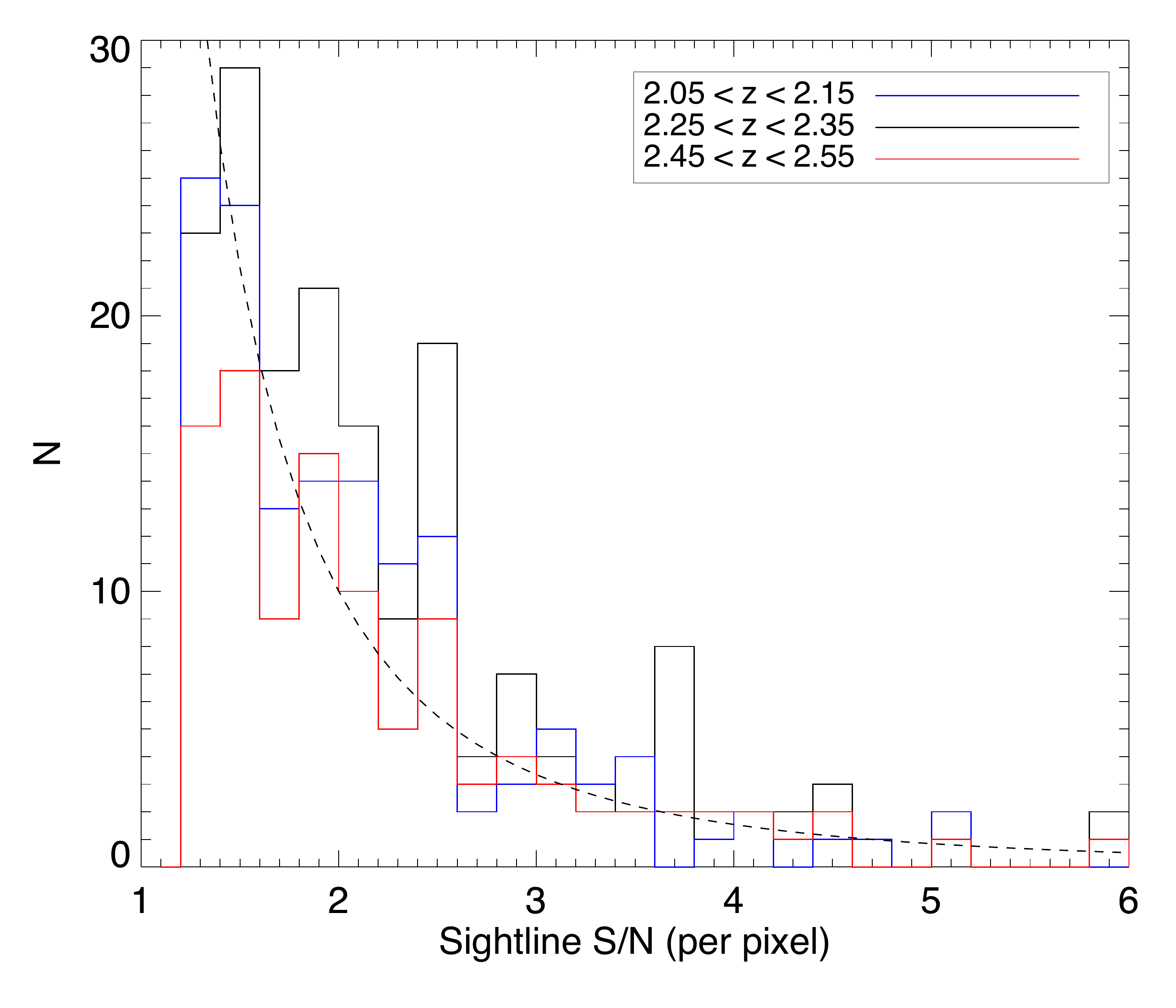}
\caption{\label{fig:snhist}
Distribution of the median sightline signal-to-noise within the \lyaf{}, evaluated at several redshift bins 
of our map volume. A small number of higher signal-to-noise sightlines have been left out by these plot axes. 
The dashed curve is a power-law with index of $-2.7$, which is a reasonable approximation for our signal-to-noise
distribution.
}
\end{figure}

Prior to \lyaf{} analysis, we first estimated the spectral signal-to-noise within the \lyaf{} of the background sources at $z>2$.
To be more specific, we evaluated the `continuum-to-noise ratio' (CNR), i.e.\ the signal-to-noise ratio relative to a 
rough initial estimate of the background source intrinsic continuum, $C$. For the LBGs, this was done as a simple power-law
extrapolation from the restframe $\lambda>1216\,\ang$ portion of the spectrum, while for the quasars we fitted principal
components to the $\lambda > 1216\,\ang$ spectrum, using templates from \citet{paris:2011}. Note that this initial continuum
  for the signal-to-noise estimation is different from that used to actually extract the \lya{} forest (Equation~\ref{eq:delta_f}, below)
 since this is much faster than the more careful mean-flux regulation used in Equation~\ref{eq:delta_f}.

We evaluated the CNR of the \lyaf{} pixels in each spectrum
over three absorption redshift ranges: $2.05<\za<2.15$, $2.15<\za<2.35$ and $2.35 < \za < 2.55$. Any high-redshift spectrum with confidence $\geq 3$ that has  
$\langle \mathrm{CNR}\rangle \geq 1.2$ over either \lyaf{} absorption redshift range was deemed useful for tomographic reconstruction.
This is an aggressive choice which incorporates nearly every background object with a confident 
redshift estimate (Figure~\ref{fig:source_zhist}), leaving out only objects that were identified primarily through 
a \lya{} emission line and therefore have negligible continua. We believe this is a reasonable approach
 since our Wiener-filtering reconstruction algorithm has noise-weighting, and \citet{lee:2014} also argued
 for such an approach in the $\dperp\gtrsim 1.5\,\hMpc$ shot-noise dominated regime which CLAMATO is in.

These position of the sightlines on the sky are shown in Figure~\ref{fig:sightlines_pos}. Note that this is a selection of 
\lyaf{} sightlines specifically probing the $2.05<\za<2.55$ \lyaf{} where we will carry out the tomographic reconstruction, 
and does not encompass all possible \lyaf{} pixels in our data set; 
some of our other pixel-based analyses may make use of different selection criteria in
position, redshift, and signal-to-noise than here.

 \begin{figure}
\includegraphics[width=0.5\textwidth]{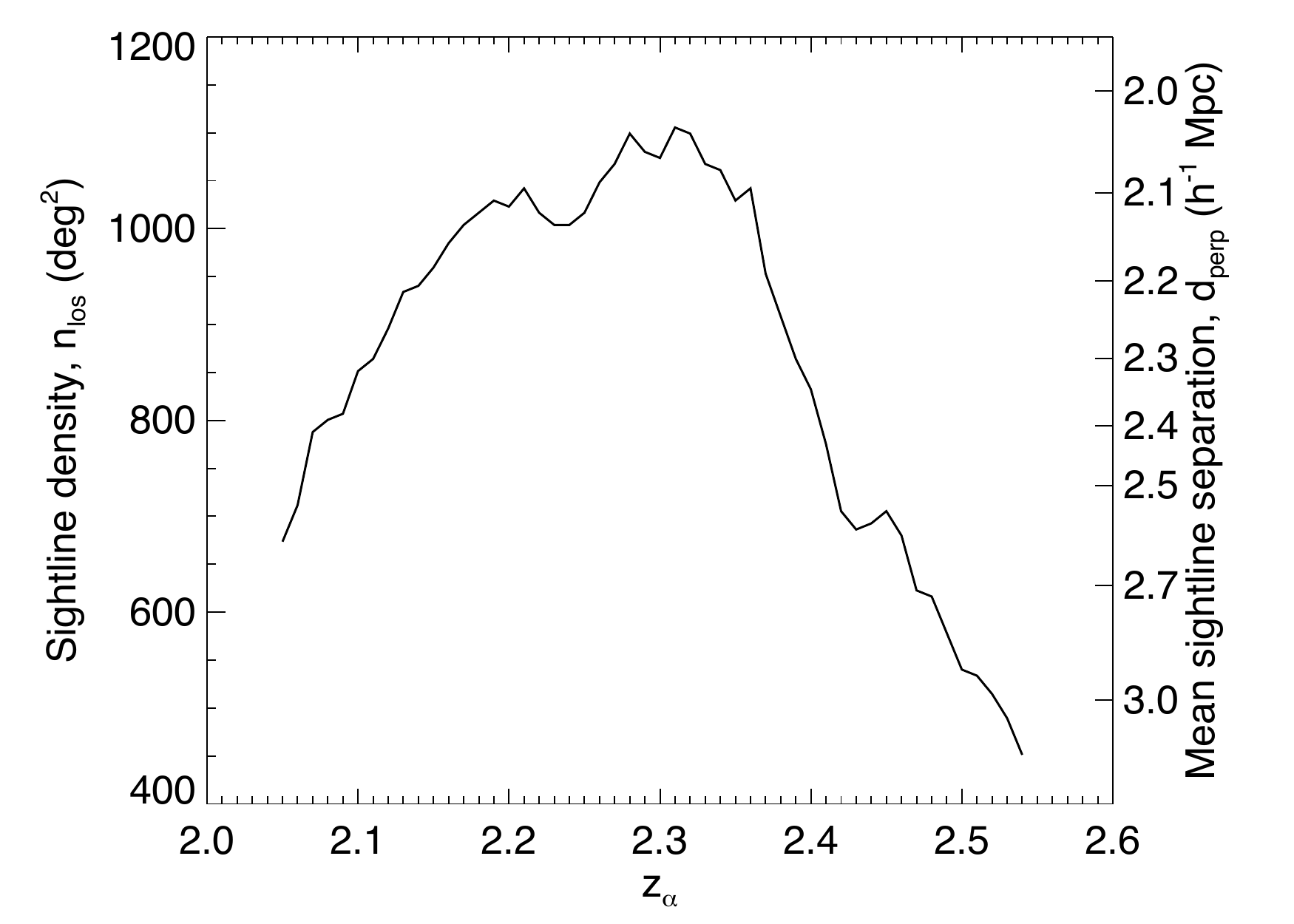} 
\caption{\label{fig:dperp_zdist}
Effective area density of Lyman-$\alpha$ forest sightlines over the redshift range of the CLAMATO tomographic
reconstruction. The right axis labels the equivalent mean separation between sightlines, $\dperp$.
The peak sightline density is $1099\,\deg^{-2}$ at $\za = 2.32$, corresponding to $\dperp = 2.04\,\hMpc$.
}
\end{figure}

There are 240 spectra within the redshift range $2.165<z_\mathrm{spec}<3.034$ (see Figure~\ref{fig:source_zhist})
that fulfilled both signal-to-noise and redshift criteria to contribute to the tomographic 
reconstruction of the foreground \lyaf{} within at least part of the redshift range $2.05<\za<2.55$. 
The distribution of estimated \lyaf{} signal-to-noise is shown in Figure~\ref{fig:snhist} at several redshifts
within our volume. A power-law with index of $-2.7$, which was adopted by \citet{krolewski:2017} and 
\citet{krolewski:2017a} is a reasonable match for this distribution.
Based on the positions of these sightlines, we defined a transverse footprint
for the tomographic reconstruction. This spans a comoving region of $26.6' \times 21.3'$  in the R.A.\ and 
declination dimensions, respectively (Figure~\ref{fig:sightlines_pos}); the center of this footprint is at
$10^h00^m41\fs23, +02\degr19'38.78''$ (J2000).
This is equivalent to a transverse comoving scale of $30\,\hMpc \times 24\,\hMpc$ 
at $\langle z \rangle=2.30$. The overall projected area density of all the sightlines that fall within the map 
footprint\footnote{For this calculation, we ignore sightlines that fall outside the map boundary (Figure~\ref{fig:sightlines_pos}), 
although they will nonetheless contribute to the tomographic reconstruction.}  is
$N_\mathrm{los}=1455\,\mathrm{deg}^{-2}$. However, due to the finite path length of \lyaf{} probed by
 each background spectrum, the differential sightline density, 
 $n_\mathrm{los} (z)= \mathrm{d}N_\mathrm{los}/dz$,
at any given redshift within the volume is somewhat lower than this.  
Averaged over the redshift range of the map, the mean sightline density is 
$\langle n_\mathrm{los}  \rangle = 866 \,\mathrm{deg}^{-2}$, equivalent to a mean sightline
separation of $\dperp = 2.35\,\hMpc$.
At the low- and high-redshift ends of the map volume ($z=[2.05,2.55]$), the effective sightline density is 
$n_\mathrm{los} = [673, 451]\,\mathrm{deg^{-2}}$, equivalent to average transverse comoving separations of 
$\dperp = [2.61, 3.18] \,\hMpc$ between sightlines (see Figure~\ref{fig:dperp_zdist}).
The effective sightline density increases towards the middle of the map redshift range, 
to a peak
density of $1099\,\mathrm{deg^{-2}}$ at  $\za = 2.32$, near the mean redshift,. This is 
equivalent to $\dperp = 2.04\,\hMpc$ comoving transverse separation.
 Note that these sightline densities are not uniformly distributed throughout the map footprint
due to shot noise as well as some background source clustering from the known galaxy overdensities at $z\sim 2.5$.

In comparison, the BOSS sightline density --- which had hitherto the best 3D sampling of the \lyaf{} --- 
is often quoted as $16\,\mathrm{deg^{-2}}$ \citep{lee:2013}, but this is in fact
the projected sightline density over all redshifts; the effective sightline density (which gives the transverse
sightline separation at a given redshift) for BOSS peaks at $9\,\mathrm{deg^{-2}}$ at $\za = 2.3$.
CLAMATO therefore represents a two order-of-magnitude increase in the sightline density probing the \lyaf{}, 
albeit over a much more limited area.

For the spectra that we want to analyze, we divide the observed spectral flux density, $f$, by the estimated continuum, $C$,
and the assumed mean \lyaf{} transmitted flux, $\langle F \rangle (z)$, at that redshift, to obtain the \lyaf{} fluctuation at each pixel:
\begin{equation}\label{eq:delta_f}
\delta_F = \frac{f}{C\;\langle F \rangle (z) } - 1.
\end{equation}
We adopt the \citet{faucher-giguere:2008a} values for $\langle F \rangle (z)$.

The intrinsic continua, $C$, of the sources is estimated differently depending on whether they are galaxies or quasars.
For the quasars, we apply PCA-based mean-flux regulation 
\citep[MF-PCA; e.g.,][]{lee:2012a, lee:2013}. Each spectrum is fitted with
a continuum template to obtain the correct shape for the intrinsic emission lines, 
which is further fitted with a linear function within the \lyaf{} region such that it yields
a mean absorption consistent with \citet{faucher-giguere:2008a}. Since the integrated forest variance 
over each $\sim 400\,\hMpc$ sightline is equivalent to only $\sim 2\%$ rms \citep[e.g.,][]{tytler:2004}
this technique allows automated continuum-fitting with $<10\,\%$ rms errors even with noisy spectra.
This technique was applied to the restframe $1041\,\ang<\lambda<1185\,\ang$ 
\lyaf{} region of the quasar spectra 
using templates from \citet{paris:2011}, masking intrinsic broad absorption where necessary. 

A similar process is applied on the galaxies, albeit assuming a fixed continuum template from
\citet{berry:2012} and adopting a more generous \lyaf{} range ($1040\,\ang<\lambda<1195\,\ang$).
We also mask $\pm7.5\,\ang$ (observed frame) around possible intrinsic absorption at 
restframe \waveion{N}{2}{1084}, \waveion{N}{1}{1134}, \waveion{C}{3}{1176},
and \ion{Si}{2} $\lambda\lambda 1190, 1193$ . We estimate that the continuum errors are approximately $\sim 10\%$ rms 
for the noisiest spectra ($\mathrm{S/N} \sim 2$ per pixel) and improving to $\sim 4\%$ rms for $\mathrm{S/N} \sim 10$ spectra \citep{lee:2012a}.

The $\delta_F$ pixel values, as well as the associated noise uncertainty, $\sigma_N$, from the pipeline, constitute the input
for the tomographic reconstruction. We have made these extracted $\delta_F$ and $\sigma_N$ pixel data 
publicly available; 
see Appendix~\ref{app:dr} for details.

The next step for the reconstruction is to define the three-dimensional comoving output grid for the map.
We choose an area spanning $26.6' \times 21.3'$ in the longitudinal and latitudinal dimensions, respectively (Figure~\ref{fig:sightlines_pos}), 
and spanning a redshift range of $2.05<z<2.55$. The angular footprint of this grid is $3.5\times$ larger than that in \citet{lee:2016},
while we have also extended the redshift range by $67\%$ from $2.20<\za<2.50$ to $2.05<\za<2.55$.
The extension to lower redshifts was because we realized that that the sightline density was higher at lower 
redshifts than originally anticipated (Figure~\ref{fig:dperp_zdist}), while we also extended to slightly higher redshifts
in order to investigate the \citet{wang:2016} galaxy cluster at $z=2.51$ despite the falling sightline density.

We adopt the simplification of a fixed Hubble parameter, $H(z)$, throughout our map volume evaluated at the
 mean redshift, 
$\langle z \rangle = 2.30$. This means that the differential comoving distance, $\mathrm{d}\chi/\mathrm{d}z$,
is constant throughout our map, such that a redshift segment of length $\delta z$ is equivalent to the
same comoving distance $\delta\chi$ everywhere in our grid.
The $26.6' \times 21.3'$ transverse footprint of the output grid therefore translates 
to a fixed transverse comoving scale of $30\,\hMpc \times 24\,\hMpc$ at all redshifts in our
 map. These approximations mean that we will have a smoothing 
 kernel (see below) that 
actually varies in size by several percent between the 
nearest and farthest ends of the map, but this simplication dramatically simplifies our map-making. 
{We also tested performing the Wiener filtering
over the pixel data transformed to $[x,y,z]$ comoving coordinates using the evolving $H(z)$ over our redshift range,
such that the sightline pixels appeared to flare outwards relative to the comoving grid.
The resulting map was found 
to have a negligible effect on the
cosmic void analysis of \citet{krolewski:2017a}, 
but breaks
the one-to-one correspondence between [RA, Dec] and the transverse $[x,y]$ coordinates of the comoving grid.
For our fiducial map, we therefore decide to use the approximation of a constant $H(z)$ over our map 
in order to preserve the one-to-one relationship between [RA, Dec] and transverse $[x,y]$, which facilitates
 comparisons with coeval galaxy positions. More detailed cosmological analyses would require the correct
 $H(z)$ to be adopted, but those would tend to directly use the pixel data rather than going through the tomographic
 map reconstruction.}

With this approximation, we thus define an output grid of $60 \times 48 \times 876$ cells each $0.5\,\hMpc$ on a side.
This cell size allows an adequate sampling of our tomographic reconstruction, which has an
effective smoothing scale of $\sim 2-3\,\hMpc$. The overall comoving volume covered by 
the output grid is thus $3.15\times 10^5\,h^{-3}\,\mathrm{Mpc}^3$. 
This is $5.4\times$ larger in comoving volume than the map described in 
\citet{lee:2016}.

For the mapmaking, we use a Wiener filtering scheme for reconstructing the sightlines 
(although see \citealt{cisewski:2014} for an alternative method).
{The basic algorithm is described in \citet{pichon:2001} and \citet{caucci:2008}}, but we use an
 implementation\footnote{\url{https://github.com/caseywstark/dachshund}} developed by \citet{stark:2015}.
 This solves for the reconstructed \lyaf{} flux
field:
\begin{equation}\label{eq:wiener}
\delta^{\mathrm{rec}}_F=\cmd\cdot (\cdd+\mathbf{N})^{-1}\cdot\delta_F,
\end{equation}
where $\cdd+\mathbf{N}$ and $\cmd$ are the data-data and map-data covariances, respectively.
This algorithm uses preconditioned conjugate gradient technique to solve the matrix inversion and matrix
multiplication steps of reconstruction.
We assumed a diagonal form for the noise covariance matrix $\mathbf{N}\equiv N_{ii}=\sigma_{N,i}^2$,
such that there only diagonal elements populated by the pixel variances $\sigma_{N,i}^2$.
However, there are a small number of spectra, primarily from bright quasars, with signal-to-noise
ratios $>10\times$ larger than the average, that could dominate the reconstruction due to the 
noise-weighting of the Wiener filter. We therefore introduced a noise floor of $\sigma_{N,i}\geq 0.2$
to the noise vector to allow a more uniform contribution from all sightlines.

We also assumeed a Gaussian covariance between any two points $r_1$ and $r_2$, such that
$\cdd=\cmd=\mathbf{C(r_1,r_2)}$ and 
\begin{equation}  \label{eq:kernel}
\mathbf{C(r_1,r_2)}=\sigma_F^2\exp\left[-\frac{(\Delta r_\parallel)^2}{2L^2_\parallel}\right]\exp\left[-\frac{(\Delta r_\perp)^2}{2L^2_\perp}\right],
\end{equation}
where $\Delta r_\parallel$ and $\Delta r_\perp$ are the distance between 
$\mathbf{r_1}$ and $\mathbf{r_2}$ along, and transverse to the line-of-sight, respectively. 
This Gaussian form was found by \citet{stark:2015a} to be a reasonable approximation 
to the true correlation function of the \lyaf.
We adopt a transverse and line-of-sight
correlation lengths of $L_\perp= 2.5\,\hMpc$ and $L_\parallel=2.0\,\hMpc$, respectively, 
as well as a normalization of $\sigma^2_F=0.05$. 
This form of covariance and parameters were determined by \citet{stark:2015a} to be approximately 
optimal for our data. Intuitively, $L_\perp$ can be thought of as set by our average sightline separation, i.e.\ $L_\perp \approx \dperp$, 
while $L_\parallel^2 \approx L_\perp^2 - \sigma_\mathrm{lsf}^2$, i.e.
it takes into account the spectral smoothing by the spectrograph, $\sigma_\mathrm{lsf}$,
to match $L_\perp$ in the line-of-sight dimension and thus provide an isotropic smoothing kernel.

We carried out the Wiener reconstruction of the map data from the 64332 input pixels with the aforementioned parameters using
the \citet{stark:2015} algorithm, with a 
stopping tolerance of $10^{-3}$ for the pre-conditioned conjugation gradient solver.
This required a run-time of approximately 
1000s using a single core of a 
Apple MacBook Pro laptop with 2.9 GHz Intel Core i5 processors and 16GB of RAM. 

{
In addition to the map itself, we have also computed the map variance associated with the Wiener reconstruction:
\begin{equation}\label{eq:variances}
\mathrm{Var}(\delta^{\mathrm{rec}}_F) = \cmd\cdot (\cdd+\mathbf{N})^{-1}\mathbf{C}_\mathrm{DM}, 
\end{equation}
where $\mathbf{C}_\mathrm{DM} \equiv \cmd^{\mathsf T}$. The variances were far more computationally
intensive to calculate than the map itself (by a factor of $N_\mathrm{pix}$), but will allow analyses to take
accunt for reconstruction uncertainties. This estimate incorporates all sources of variances in the map, including pixel noise, finite skewer
sampling, and intrinsic variance of the \lyaf. We will further discuss these in Section~\ref{sec:discussion}.
}

The resulting map and standard deviations is publicly available for download as a binary file; 
see Appendix~\ref{app:dr} for details.

\section{Results}\label{sec:results}

\begin{figure*}\centering
\includegraphics[width=\textwidth, clip=true, trim=155 5050 212 30]{slice_yz_Lpar2p0_Lperp2p5_v4_sm2p0_skewers_skymap.pdf}
\includegraphics[width=\textwidth, clip=true, trim=155 2340 212 260]{slice_yz_Lpar2p0_Lperp2p5_v4_sm2p0_skewers_skymap.pdf}
\caption{\label{fig:slicemaps}
Wiener-filtered tomographic reconstructions of the \lyaf{} absorption field, $\delta_F^{\mathrm{rec}}$,
 at $2.05<\za<2.55$ from the
current CLAMATO data (color map), shown after smoothing with an isotropic $R=2\,\hMpc$ Gaussian kernel. 
Each color panel shows the absorption projected over a $2\,\hMpc$ R.A. slice, 
the position of which is denoted by the shaded region in the subpanels to the left that also show the sightline
positions on the sky as red dots. The color convention for the absorption is such that red denotes overdensities 
while blue denotes underdensities. White horizontal lines denote the sightline coverage,
while symbols mark the location of known foreground galaxy redshifts: downwards triangles from MOSDEF, 
upwards triangles from ZFIRE, squares from VUDS, diamonds from zCOSMOS-Deep, and circles from CLAMATO. 
The large black stars indicate the reported central positions of the galaxy overdensities at $z=[2.10, 2.44, 2.47, 2.51]$.
This sequence is continued in Figure~\ref{fig:slicemaps2}.
}
\end{figure*}

\begin{figure*}
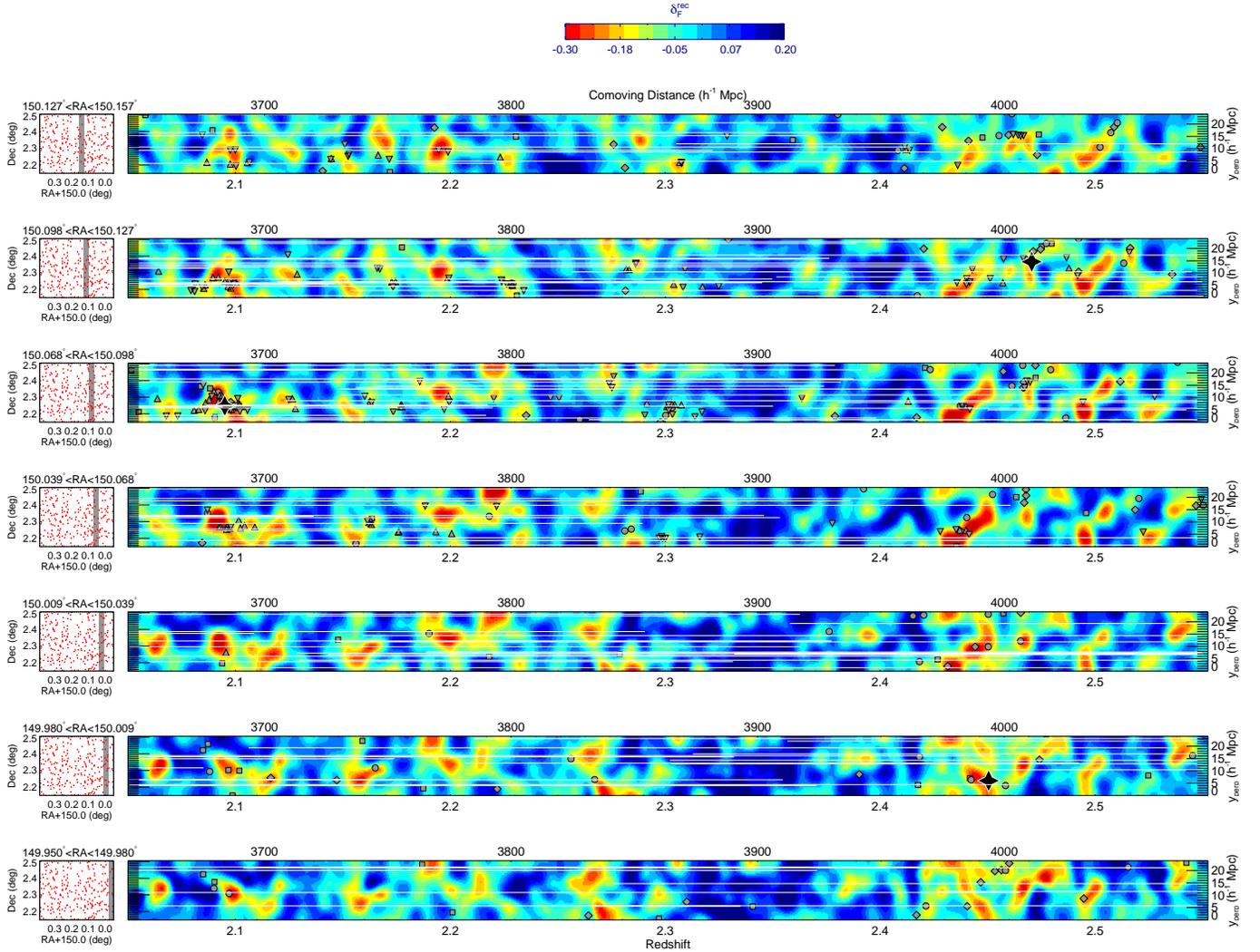
\centering
\includegraphics[width=\textwidth, clip=true, trim=155 5050 212 30]{slice_yz_Lpar2p0_Lperp2p5_v4_sm2p0_skewers_skymap.pdf}
\includegraphics[width=\textwidth, clip=true, trim=155 0 212 2940]{slice_yz_Lpar2p0_Lperp2p5_v4_sm2p0_skewers_skymap.pdf}
\caption{\label{fig:slicemaps2}
Continued from Figure~\ref{fig:slicemaps}.
}
\end{figure*}

\begin{figure*}[th]\centering
\includegraphics[width=0.65\textwidth,clip=true, trim=0 90 0 160]{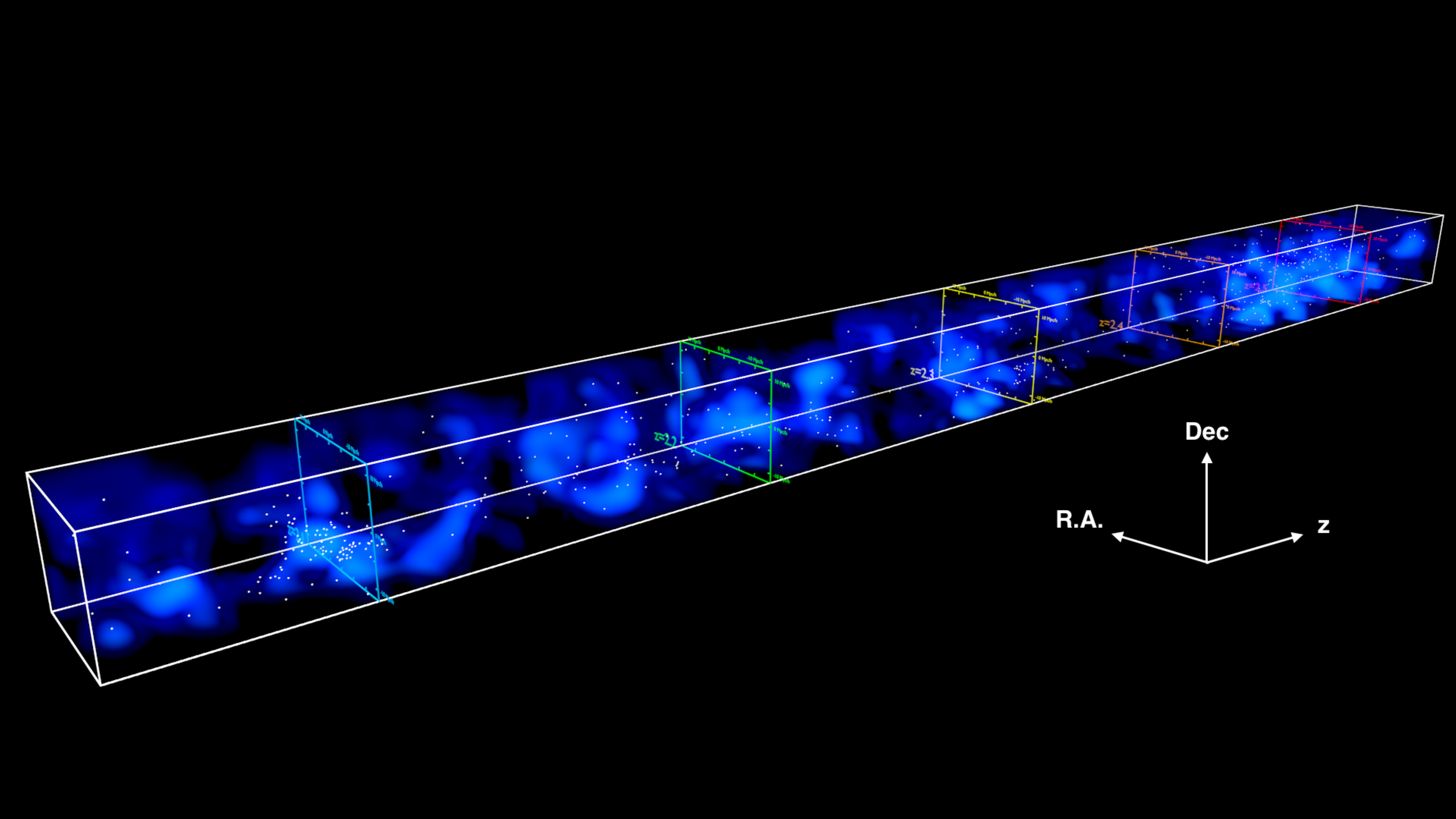}
\caption{\label{fig:screenshot}
Still image from our 3D video visualization of the CLAMATO reconstructed absorption map 
(smoothed with a $R=2\,\hMpc$ Gaussian kernel), where the absorption is indicated 
by the blue transparency. Foreground galaxy redshift positions are denoted by the yellow dots, while
the triad (not present in the video) indicates the directions of increasing R.A., declination, and redshift. 
The video is available in the online journal. 
Alternatively, the YouTube version (\url{https://youtu.be/QGtXi7P4u4g}) offers a virtual-reality 
option when viewed with a smartphone and Google Cardboard-compatible headset.
}
\end{figure*}

{
In Figure~\ref{fig:slicemaps} we show a slice visualization of the resulting tomographic map, 
where we have divided the three-dimensional volume into
projected slices over the longitudinal (i.e.\ R.A.) direction with thicknesses of $2\,\hMpc$. 
The $x$-axis of each slice therefore denotes the redshift or line-of-sight dimension, while the $y$-axes
are along the declination or latitudinal dimension in the plane of the sky. For clarity, we have found it useful
to further smooth the map with a Gaussian kernel, in this case with 
standard deviation $R=2\,\hMpc$. 
For comparison, we have also overplotted the positions of 552 known coeval spectroscopic redshifts that overlap
our map volume, which are primarily from  
zCOSMOS-Deep \citep{lilly:2007} and VUDS \citep{le-fevre:2015}, but also from publicly-released redshifts 
such as MOSDEF \citep{kriek:2015} and ZFIRE \citep{nanayakkara:2016}. We also included the positions
of our own CLAMATO galaxies that fell within the foreground map volume. For the galaxies with spectroscopic redshifts
from more than one survey, we used the redshift estimates in the following order of descending priority:
 MOSDEF, ZFIRE, CLAMATO, VUDS, then zCOSMOS-Deep. In addition to the two-dimensional visualization, 
 we have also created a video visualization (Figure~\ref{fig:screenshot}) as well as a manipulable interactive
 figure for the online journal (Figure~\ref{fig:x3d}); more details on these 3D visualizations are given in Appendix~\ref{app:viz}.}

While there are multiple science analyses in preparation based on the CLAMATO data presented in this paper, 
here we qualitatively discuss the more noteable features apparent in the
 tomographic \lyaf{} absorption map described in the previous section.

\begin{figure*}\centering
\includegraphics[width=0.65\textwidth,clip=true, trim=0 0 0 0]{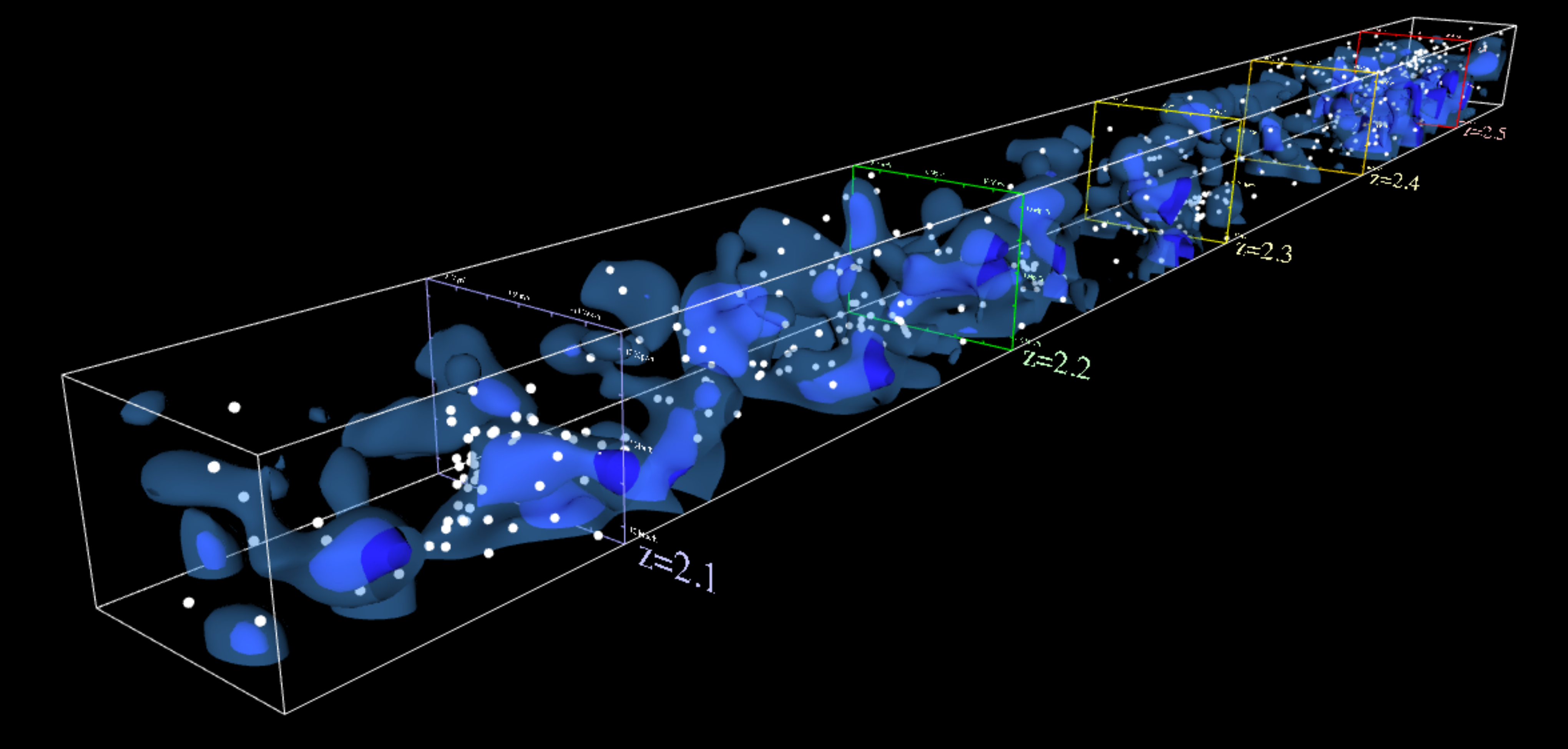}
\caption{\label{fig:x3d}
Three-dimensional rendering of the CLAMATO tomographic map, showing two isodensity contours at $\deltarec = -0.08$
and $\deltarec=-0.18$, along with coeval galaxy positions shown as dots. This figure is available online as an
interactive figure (\url{http://www.mpia-hd.mpg.de/homes/tmueller/projects/clamato/map2017.html}) --- it requires a load-time of several minutes. 
By left-clicking and moving the mouse, 
the viewpoint can be rotated, while the right-mouse button or scroll wheel can be used to zoom in or out; 
double left-clicking at any point in the map focuses the viewpoint there.
The buttons labeled `Isosurface: -0.08' and `-0.18' toggles the respective isodensity surfaces on and off.
The 'Reset View' button restores the figure to its default state and perspective.
}
\end{figure*}

 \subsection{Large-Scale Structure Features}
 
  In all these visualizations, the IGM absorption and the coeval galaxies generally appear to
 trace the same structures.  However, the foreground galaxy redshifts are incomplete across our volume, 
 and several of the spectroscopic surveys (i.e. MOSFIRE and ZFIRE) target only a limited sub-field within the central
 portion of the CLAMATO footprint (see Figure~\ref{fig:survey_fields}).
 It would therefore be challenging to construct a uniform density map from the galaxy redshifts, whereas
 the tomographic map delivers a more detailed view of large-scale structure in that volume.
   In upcoming papers, we will investigate the 
 relationship between galaxy properties and their density environment, assuming that the \lyaf{} absorption
 traces large-scale structure. There will be alternative analyses using different formalisms: a direct comparison of 
 galaxy positions with the local IGM absorption, and a pixel-level cross-correlation analysis 
 analogous to those carried
 out in BOSS for quasars and damped \lya{} absorbers \citep{font-ribera:2012, font-ribera:2013}.
 In both cases, we will aim to carry out the investigations as a function of galaxy properties fitted from their
 spectral energy distributions. A recent study \citep{sorini:2017} has also argued that the precise shape of the
  galaxy-forest cross-correlation on $\sim 1\,\hMpc$ scales could be used to place constraints on galaxy feedback, 
  which will make feedback models another parameter space we could investigate.
 
With reference to the galaxy redshifts alone, 
 an apparent lack of galaxies at any point in space in these visualizations do not
 necessary imply a true absence of galaxies due to the incompleteness of the galaxy surveys. 
 But in the IGM map, we clearly see large coherent underdensities, with a notably striking underdensity at $z\approx 2.35$
 appearing to extend $>10\,\hMpc$ along both the transverse and line-of-sight dimensions. 
 These underdensities also appear to be devoid of galaxies and therefore are likely to be true cosmic voids.
 A detailed analysis of the cosmic voids 
 in the CLAMATO map, which are by far the most distant such objects ever found,
  is presented in a companion paper \citep{krolewski:2017a}.

Conversely, we see excess absorption corresponding to multiple galaxy overdensities that have been 
identified through other methods.
In particular, we clearly see the extended \lya{} absorption signature from 
the $z\approx 2.5$ overdensity comprised of 
the $z=2.44$ protocluster \citep{diener:2015, chiang:2015}, $z=2.47$ protocluster \citep{casey:2015}, and
X-ray detected $z=2.51$ cluster  \citep{wang:2016}. 
In the 3D visualizations (Figure~\ref{fig:screenshot} and \ref{fig:x3d}), we see that these structures appear to form a 
giant interconnected structure extending roughly from $2.44 < z < 2.52$ with a complex topology.
Another known overdensity seen in our map is the
$z=2.095$ galaxy protocluster initially identified through the ZFOURGE medium-band 
photometric redshift survey \citep{spitler:2012}
and subsequently confirmed with NIR spectroscopy \citep{nanayakkara:2016}.
In an upcoming paper we will analyze the properties of these overdensities in conjunction with large-volume hydrodynamical simulations, 
although further CLAMATO data will be required in order to fully map out the extent of these overdensities since
they overfill our current map boundaries. In particular, we are interested in the fate of the $z=2.44-2.51$ system
of overdensities: \citet{wang:2016} have argued that the $z=2.51$ overdensity, in itself, 
might collapse into a $2\times 10^{15}\,\mathrm{M}_\odot$ galaxy cluster at late times, i.e.\ it might be
fated to become one of the most massive clusters in the known universe. With the detailed large-scale structure
information from IGM tomography, we aim to carry out a detailed investigation into the evolution of these
structures, especially using
constrained realization techniques \citep[e.g.,][]{jasche:2015, wang:2014}.

\section{Quality Assessment}\label{sec:discussion}

\subsection{Wiener-based Map Variance}

{Firstly, the variances estimated from the Wiener filtering algorithm (Equation~\ref{eq:variances}) provide the most obvious approach to
quantify the fidelity of the map. We find that mean value for the estimated variance is
$\langle\mathrm{Var}(\delta^{\mathrm{rec}}_F) \rangle = 0.0219$ within the $L=0.5\,\hMpc$ 
volume elements (``voxels'') in our reconstruction. 
This variance includes contributions from the intrinsic variance of the \lyaf as well as pixel noise and finite sightline sampling.
In order to estimate the uncertainties caused by noise and sampling variance contributions, we therefore need to subtract the intrinsic variance.
We do this using the simulated fluxes from the $L=256\,\hMpc$ N-body simulations described in 
\citet{stark:2015, stark:2015a}, which we binned to the same voxel size and smoothed with a Gaussian kernel of the same
size as applied to the real map (Equation~\ref{eq:kernel}).
The intrinsic \lyaf{} variance estimated from the simulation is $\mathrm{Var}^\mathrm{intr}(\delta_F)=0.00654$.}

{We can thus proceed to define the global signal-to-noise ratio as:
\begin{equation}\label{eq:snwiener}
\mathrm{S/N}^\mathrm{wiener}\equiv \sqrt{\frac{  \mathrm{Var}^{\mathrm{intr}}(\delta_F) }
{  \langle\mathrm{Var}(\delta^{\mathrm{rec}}_F) \rangle  -  \mathrm{Var}^{\mathrm{intr}}(\delta_F) } } ,
\end{equation}
i.e.\ the square-rooted ratio of the intrinsic variance of the \lyaf\ (the signal) compared with the variance contributions from the pixel noise
and finite sightline sampling. For our current map, we find $\mathrm{S/N}^\mathrm{wiener} \approx 0.65$ per 
individual $0.5\,\hMpc$ voxel.
Over larger scales used for most analyses, the signal-to-noise is commensurately improved as the square-root of the number of
pixels being averaged over. For example, over top-hat kernels of $[2, 3, 4]\,\hMpc$, the signal-to-noise
would on average be improved to $\mathrm{S/N}^\mathrm{wiener} \approx [1.8, 3.4, 5.2]$, respectively.    }

\subsection{Comparison with Forecasts of \citet{lee:2014}}
{In \citet{lee:2014}, we made predictions for the quality of IGM tomographic maps based on various observational scenarios.
We now compare our actual data with the earlier forecasts.
In \citet{lee:2014}, we defined the 
following quantity\footnote{This quantity was
denoted as S/N$_{\epsilon}$ in \citet{lee:2014}, but here we rename it to avoid confusion with the quantity defined in Equation~\ref{eq:snwiener}.} based on the deviation of mock tomographic reconstructions with respect to the true underlying
flux in the simulations:}
\begin{equation}
S_\epsilon = \sqrt{\frac{\mathrm{Var}(\delta^{\mathrm{true}}_F)}{\mathrm{Var}(\deltarec - \delta^{\mathrm{true}}_F)}},
\end{equation} 
where $\delta^{\mathrm{true}}_F$ is the true \lyaf{} flux field from the simulation and 
$\deltarec$ is the tomographic reconstruction
of mock data from the same volume.
For this purpose, 
we use the aforementioned $L=256\,\hMpc$ N-body simulations. 
We first divide up the simulation volume into $32\,\hMpc \times 32\,\hMpc \times 256\,\hMpc$
chunks to approximate the elongated CLAMATO survey geometry, randomly drawing \lyaf{} absorption
 skewers with a mean sightline separation of $\dperp = 2.5\,\hMpc$, and then adding Gaussian random
 noise to each sightline's pixels, consistent with the signal-to-noise distribution of the CLAMATO sightlines.
We also introduced a random continuum error to each sightline based on the 
 sightline signal-to-noise: we assumed a inverse relationship between the signal-to-noise and continuum error, 
 such that, e.g. a $\mathrm{S/N} = 2$ sightline gets 12\% continuum error while a  $\mathrm{S/N} = 10$ sightline gets only 
 $3.5\%$ continuum error (for more details, see \citealt{krolewski:2017a}).
 The sightlines from each mock survey are then Wiener-reconstructed the same way as the CLAMATO data.
 
 
 Following the prescription from \citet{lee:2014}, we then Gaussian-smooth both the true and reconstructed flux fields
 with a $R=4\,\hMpc$ kernel (i.e.\ a smoothing kernel $1.4\,\times$ the mean sightline separation) before calculating the 
 simulation-estimated signal-to-noise. For CLAMATO, we find $S_\epsilon = 2.26$ after averaging over 64 mock survey volumes.
 This is a slightly conservative estimate since the $\dperp=2.5\,\hMpc$ sightline separation assumed in the mocks is sparser than the average
 $\dperp=2.37\,\hMpc$ over our entire map, but it is within the $S_\epsilon \sim 2-2.5$ range of what \citet{lee:2014} considered a good reconstruction quality. We also cross-checked
 this with the analytic method for calculating $S_\epsilon$ (Equation 18 in \citealt{lee:2014}), 
 which takes as input the sightline signal-to-noise distribution. 
 This calculation yields  $S_\epsilon = 2.30$,  which is in good agreement with the estimate from the
 mock reconstructions\footnote{To assist in planning of future IGM tomography surveys, we have made the
 analytic code publicy available under an MIT license: \url{https://github.com/kheegan/tomo_mapsn} and archived Version 1 on Zenodo \citep{kheegan_2018_1293048}.}. 
 This suggests that the
 analytic formalism would be a useful tool for forecasting future IGM tomography surveys to provide signal-to-noise
 estimates relative to CLAMATO.
 
 However, in retrospect we now find the forecasts from \citet{lee:2014} to be optimistic compared to what we have been obtaining
 with CLAMATO. In particular, the forecasted area density of LBGs at fixed magnitude is considerably lower than what we observe.
 \citet{lee:2014}, for example, projected a sightline density of $660\,\deg^{-2}$ at a magnitude limit of $g\leq 24.2$ whereas we have
 the equivalent of $344\,\deg^{-2}$ at the same limit.
 {This shortfall is apparent not just within the present CLAMATO data, but also when looking at all possible targets with 
 the appropriate brightness and photometric redshifts across the
 full COSMOS field based on the \citet{laigle:2016} catalog}. We believe this is a genuine discrepancy and
 attribute it to the likely combination of several factors: (i) a mismatch between the $g$ filter assumed in  
 \citet{lee:2014} and the different filter set of \citet{reddy:2008}, whose luminosity function was used to estimate sightline
 availabilty, (ii) uncertainties in the luminosity function, whose error bars are a factor of 2 or 3 at the bright end. Due to 
 the steep slope at the bright end of the luminosity function, even small discrepancies could translate to large differences in number count.
 
 The scaling of spectral signal-to-noise with exposure time in \citet{lee:2014} was also found to be too optimistic. The older
 paper assumed, for example, that a 4hr exposure with the VLT (equivalent to 2.6hrs on with the larger Keck telescope) would yield S/N=4 per angstrom on a $g=24.0$. We find, on the other hand, that a comparable exposure time yields only S/N$\approx$3 per angstrom 
 (with a considerable scatter) on 
 a similar source magnitude. This is most likely due to the fact that \citet{lee:2014} assumed that the star-forming (and hence UV-emitting) regions of the background galaxies are point sources, 
 whereas real LBGs are sufficiently extended as to increase the amount of sky background
 noise beyond that assumed by \citet{lee:2014}.
  
 In the CLAMATO observations, we made up for these shortfalls by filling our slitmasks with targets even if they
 fall below our nominal survey limit, and then being aggressive in incorporating low signal-to-noise spectra into our
 tomographic reconstruction. \citet{lee:2014} calculated that adding more, low signal-to-noise, spectra is a viable 
 survey strategy to boost the tomographic map signal-to-noise in the $\dperp\gtrsim 1\,\hMpc$ shot-noise dominated
 scales which CLAMATO is probing. We have also reobserved many fields within our footprint, both to obtain 
 additional integration times or with redesigned slitmasks as new targeting information became available. Our
 sightline coverage is therefore more homogeneous than if we had pursued a single-pass strategy with fixed
 exposure time, and even then there are gaps in the footprint that we were not able to fill even 
 after 10hrs of integration (see Table~\ref{tab:slitmasks}). 
 
 We were also likely helped by the presence of the overdensities at $z\sim 2.44-2.51$, which provided additional sightlines 
 for the $\za<2.4$ map region in their foreground. We therefore expect our mean sightline separation to increase from the current
 $\dperp =2.37\, \hMpc$ as the survey footprint extends into the rest of the COSMOS field.

\section{Conclusion}
In this paper, we have described the first data release of the CLAMATO Survey, the first systematic attempt
at implementing 3D \lyaf{} reconstruction on several-Mpc scales using high area densities 
($\sim 1000\,\deg^{-2} $) of background LBG and quasar spectra.

With Keck-I LRIS observations of 23 multi-object slitmasks over $0.157\mathrm{deg}^2$ in the COSMOS field, 
we obtained 293 spectra with confident redshifts, 
of which 240 were at the right redshift and had sufficient signal-to-noise to used as background sources
probing the $2.05< \za  <2.55$ \lyaf{}.
The average transverse separation between these sightlines is only $\dperp = 2.35\,\hMpc$.
We used these spectra to create a three-dimensional tomographic map of the IGM absorption at these redshifts, 
which has comoving dimensions of $30\,\hMpc \times 24\,\hMpc \times 438\,\hMpc 
\simeq 3.15\times 10^5\,h^{-3}\,\mathrm{Mpc}^3$. We have made all the catalogs, spectra, pixel data, and 
reconstructed maps publicly available (see Appendix~\ref{app:dr}). 

By eye, the CLAMATO absorption map appears to trace similar structures as the coeval galaxies with known 
spectroscopic redshifts within the COSMOS field, and also reveals large extended structures associated with
several known galaxy overdensities in the field. There are also clear underdensities that are also devoid of galaxies 
and hence correspond to cosmic voids \citep{krolewski:2017a}.
Multiple science analyses are now ongoing on this data, including measuring the cross-correlation between
the \lyaf{} and coeval galaxies, studying galaxy properties as a function of IGM environment, and analysis
of the protoclusters in the volume.

Over the next few years, we hope to expand the CLAMATO map to at least $0.5\,\deg^2$, which will achieve a cosmological
volume of $10^6\,h^{-3}\,\mathrm{Mpc}^3$. This will give full coverage of the large overdensities we currently see in 
the map, and cover $\sim 1200$ coeval galaxies, which would offer sufficient statistical power for comparative studies
of their properties as a function of IGM environment. For cosmology, preliminary estimates 
suggest that the full CLAMATO survey will have comparable numbers of unique 3D \lyaf{} pixel-pairs
at several-Mpc separations as the 1D pixel-pairs at similar scales used in the BOSS DR9 one-dimensional 
forest flux power spectrum measurement \citep{palanque-delabrouille:2013a}. This could allow interesting complementary constraints on cosmological parameters
such as the sum of neutrino masses and the
curvature of the primordial density fluctuation power spectrum. 
Another interesting measurement that could be attempted with the CLAMATO data is the weak-lensing of the 
\lyaf{} \citep{croft:2017}, which uses the gravitational deflection of the $z\sim 2-3$ \lyaf{} to probe the $z\sim1$ 
matter field, which is at a higher redshift than currently probed by 
galaxy cosmic shear weak lensing measurements.
Based on the estimates from \citet{metcalf:2017}, the $0.5\,\deg^2$ CLAMATO survey should be able to detect
\lyaf{} weak lensing at $\sim 6\sigma$ confidence over a foreground redshift range of $\Delta z =0.5$ --- this
signal should be even stronger in cross-correlation with the rich photometric and spectroscopic redshift information
 available
for foreground galaxies in the COSMOS field.

Prior to the CLAMATO survey, \lyaf{} tomography was considered only feasible with future 30+m class 
telescopes. We have now, however, shown that the technique is in fact accessible to 8-10m class telescopes, 
enabling the mapping of the $z\sim 2-3$ IGM absorption on comoving scales of $\sim 2-3\,\hMpc$.
This demonstration is particularly exciting in the context of the various wide-field spectroscopic facilities on 8-10m
telescopes that are 
either being built, e.g. the Prime Focus Spectrograph (PFS) on the 8.2m Subaru Telescope \citep{sugai:2015},
or in various stages of planning and discussion, e.g. the 11.25m Maunakea Spectroscopic 
Explorer \citep[MSE,][]{mcconnachie:2016a}.
These facilities, which offer multiplex factors of several thousand over $\sim 1\deg^2$ fields-of-view, should be able to carry out IGM tomography over much larger areas of tens or hundreds of square
degrees, enabling new science cases at $z\sim 2-3$ with unprecedented statistical power.

Looking further into the 2020s, 30+m class facilities \citep{evans:2012, skidmore:2015} would be required to push the spatial resolution of 
IGM tomography to comoving scales of $1\,\hMpc$ and below. As \citet{lee:2014} calculated, not only do the density
of background sightlines need to increase, but the minimum pixel signal-to-noise also needs to be improved 
as these scales are no longer in the shot-noise dominated regime. The amount of photons that needed to be
collected in this regime increases exponentially as smaller mapping scales are desired, necessitating 30+m
apertures.

Finally, the 2030s could see a dedicated ``hyper-multiplexed'' ($>10^4$ multiplex) wide-field spectroscopic 
facility such as the 
Billion Object Apparatus \citep[BOA,][]{dodelson:2016} on a 10m-class survey telescope. 
While BOA will not represent a large leap in collecting area compared to Subaru-PFS or MSE, its hyper-multiplexing
will enable it to simultaneously carry out an all-sky galaxy redshift survey out to $z\sim 1.5-2$, 
and at the same time carry out 
an IGM tomography survey with similar parameters as CLAMATO, but over $\sim 10000\,\deg^2$.
The goal of such a survey would be to map all cosmological linear modes out at $0\lesssim z \lesssim 3$ in order
to push cosmological parameter constraints beyond the LSST and DESI
`Stage IV' limits.

CLAMATO, and its pioneering analyses, will be needed to pave the path for these ambitious projects of the future.

\acknowledgements{
We thank Suk Sien Tie for pointing out some bugs in our initial data release.
 K.G.L. acknowledges support for this
work by NASA through Hubble Fellowship grant HF2-51361
awarded by the Space Telescope Science Institute, which is
operated by the Association of Universities for Research in
Astronomy, Inc., for NASA, under contract NAS5-26555.
  We are also grateful to the entire COSMOS collaboration for their assistance and helpful discussions.
The data presented herein were obtained at the W.M. Keck Observatory, 
which is operated as a scientific partnership among the California Institute of Technology, 
the University of California and the National Aeronautics and Space Administration (NASA). 
The Observatory was made possible by the generous financial support of the W.M. Keck Foundation.
  The authors also wish to recognize and acknowledge the very significant cultural role and reverence that the summit of Maunakea has always had within the indigenous Hawai'ian community.  We are most fortunate to have the opportunity to conduct observations from this mountain.
  }

\appendix

\section{Data Release}\label{app:dr}
We have made the first data release of the Keck-CLAMATO data publicly available on Zenodo 
\dataset[doi:10.5281/zenodo.1292459]{https://doi.org/10.5281/zenodo.1292459}. 
These include the reduced spectra, continuum-normalized
 \lyaf{} pixels used as the input for the tomographic reconstruction, and the tomographic map of the $2.05<z<2.55$ IGM.  
 
 The 437 blue and 185 red reduced LRIS spectra are in FITS format, each with the following extensions: 
 \begin{itemize}\setlength\itemsep{0.1pt}
 \item {HDU0:} Object spectral flux density, in units of $10^{-17} \mathrm{ergs\,s^{-1}\,cm^{-2}\,\ang^{-1}}$
\item {HDU1:} Noise standard deviation
\item {HDU2} Pixel Wavelengths in angstroms
 \end{itemize}
 On the data webpage, we have provided an ASCII catalog that contains the information in Table~\ref{tab:tomo_obj} 
 as well as the corresponding filenames of
 the blue and red spectra for each object. We note that the spectrophotometry, especially in the red, might be unreliable.
 
We also provided a binary file with the intermediate product of 64332 concatenated \lyaf{} pixels (Equation~\ref{eq:delta_f})
at $2.05<\za<2.55$ from 240 background sources that satisfy our redshift and signal-to-noise criteria.
This file includes the $\delta_f$ values and associated pixel noise, as a function
of the $[x,y,z]$ positions relative to our tomographic map grid. The $x$ and $y$ 
coordinates correspond to transverse comoving distance along R.A. and Declination, respectively, 
with the origins at
$[\alpha_0, \delta_0] = [9^h 59^m 47\fs999, +02\degr9'0.00"]$ (J2000) or $[\alpha_0, \delta_0] = [149.9500\degr, 2.1500\degr]$, while
$z$ corresponds to line-of-sight comoving distance relative to the origin redshift of $\za=2.05$. 
As described in Section~\ref{sec:tomo}, we adopt a fixed conversion between comoving distance and redshift,
 evaluated at our median map redshift of $\langle z \rangle = 2.30$. With our choice of cosmology, this yields 
 $\chi = 3874.867\,\hMpc$ and
 $\mathrm{d}\chi/\mathrm{d}z = 871.627\,\hMpc$.
This intermediate binary file is the primary input used for the Wiener reconstruction algorithm to create the tomographic map.

The primary products are the binary files containing the IGM tomographic map, which spans comoving dimensions of 
$30\,\hMpc \times 24\,\hMpc \times 438\,\hMpc$ in the $[x,y,z]$ dimensions, respectively, with
 binning in units of 0.5\,\hMpc. The standard deviations ($\mathrm{Var}^{1/2}(\delta^{\mathrm{rec}}_F)$) of the reconstruction 
 (Equation~\ref{eq:variances}) are provided in a separate file with the same spatial binning and format. 
 The conversion of the map coordinates back to R.A., Declination and redshift can be carried
 out with with the aforementioned $\chi$ and $\mathrm{d}\chi/\mathrm{d}z$ values. We provide both the direct tomographic
 reconstruction of the data, as well as a version which has been Gaussian-smoothed with a $\sigma=2\,\hMpc$ kernel; the latter
 is the version shown in the visualizations in Figures~\ref{fig:screenshot} and \ref{fig:x3d}.
 
 \section{Three-Dimensional Visualizations}\label{app:viz}

We used the Blender 
software\footnote{\url{https://www.blender.org/}} to create three-dimensional video of the
tomographic maps presented in this paper.
While it is not a commonly-used tool for scientific visualization, Blender
offers superior scene
design and camera handling to most scientific visualization packages.
Because our tomographic map consists only of scalar values, we can apply direct volume 
rendering such that each density value is mapped to a particular color and opacity
value via a transfer function. 
To accomplish this, we make use of Blender's internal render engine where scalar values on a Cartesian grid 
can be represented as voxel data and the transfer function can be defined using a color ramp.
The galaxies are represented by small spheres 
which all 
have the same size --- in the future, we will aim to incorporate the morphologies and colors of the individual 
galaxies into the visualization.
We have also created a 360-degree video that is compatible with the YouTube 360 Video API or planetarium projectors. 
As the internal render engine in Blender has no full-sky camera, we have to render six orthogonal 
camera images per frame
for each camera position, with each camera's field-of-view set to $90\mathrm{deg}\times 90\mathrm{deg}$.
All six images are then assembled into a so-called cube-map image which is subsequently mapped to a equirectangular projection as needed for $360\mathrm{deg}$ videos by means of a small OpenGL program.

This video can viewed in the online version of Figure~\ref{fig:screenshot}, while a spherically-projected version has been uploaded to YouTube\footnote{\url{https://youtu.be/QGtXi7P4u4g}}
that can be displayed with their 360 Video API, which allows the viewer to 
pan the viewing angle on most common web browsers by clicking and dragging with a mouse or trackpad. 
For users viewing the video
with the Android or iOS
YouTube smartphone application, this also exploits smartphone gyroscopes and accelerometers to offer a limited virtual-reality (VR) experience
in conjunction with affordable Google Cardboard-compatible stereoscopic headsets. The viewer can turn his or her head 
to 
vary the camera viewpoint over the three rotational degrees of freedom (yaw, roll, and pitch) but not the three translational degrees.  

Figure 12 shows another alternative method of viewing the 3D map: an interactive online X3D figure \citep{vogt:2016}, 
which allows readers of the online version to pan and zoom the map 
viewpoint\footnote{\url{http://www.mpia-hd.mpg.de/homes/tmueller/projects/clamato/map2017.html}} within their web browser. 
The rendering capabilities of the X3D pathway is somewhat more
limited than the Blender software used to create the video, in that it cannot render a complicated transfer function
of the map opacity, so we have only chosen to show two iso-density contours at
 $\deltarec=-0.08$ and $\deltarec = -0.18$ with the former as a transparent blue layer while the latter is opaque, 
 along with the positions of the coeval galaxies.

\bibliographystyle{aasjournal}

\bibliography{lyaf_kg,apj-jour,lss_galaxies,my_papers}

\end{document}